\RequirePackage{scrlfile}
\ReplacePackage{scrpage2}{scrlayer-scrpage}
\documentclass[zpreprint,british]{./LaTeX/zeus/zeus_Kpaper}
\usepackage{./LaTeX/zeus/zeusdet_units}

\DeclareOldFontCommand{\bf}{\normalfont\bfseries}{\mathbf} 
\DeclareOldFontCommand{\it}{\normalfont\itshape}{\mathit}
\usepackage{./LaTeX/styles/lineno}
\chardef\usc=95
\chardef\til=126
\catcode`\@=11 
\DeclareRobustCommand\xdotspace{\futurelet\@let@token\@xdotspace}
\def\@xdotspace{%
  \ifx\@let@token.\else
  \ifx\@let@token\bgroup.\else
  \ifx\@let@token\egroup.\else
  \ifx\@let@token\/.\else
  \ifx\@let@token\ .\else
  \ifx\@let@token~.\else
  \ifx\@let@token!.\else
  \ifx\@let@token,.\else
  \ifx\@let@token:.\else
  \ifx\@let@token;.\else
  \ifx\@let@token?.\else
  \ifx\@let@token/.\else
  \ifx\@let@token'.\else
  \ifx\@let@token).\else
  \ifx\@let@token-.\else
  \ifx\@let@token\@xobeysp.\else
  \ifx\@let@token\space.\else
  \ifx\@let@token\@sptoken.\else
   .\space
   \fi\fi\fi\fi\fi\fi\fi\fi\fi\fi\fi\fi\fi\fi\fi\fi\fi\fi}
\catcode`\@=12 

\newcommand{\stru}[2]{%
   \relax\ifmmode\hbox{\vrule height#1 depth#2 width0pt}%
   \else\vrule height#1 depth#2 width0pt\fi}

\newcommand{\Ronum}[1]{\uppercase\expandafter{\romannumeral#1}}
\newcommand{\ronum}[1]{\expandafter{\romannumeral#1}}
\DeclareRobustCommand{\LaTeXZ}{%
  \LaTeX\kern-.05em4\kern-.1em
  {\raisebox{-0.2ex}{$\scriptstyle\text{ZEUS}$}}\xspace}

\newcommand{\slashfrac}[2]{%
  \raisebox{0.5ex}{\ensuremath #1}\kern-0.12em/\kern-0.08em
  \raisebox{-.8ex}{\ensuremath #2}}

\newcommand{\sqr}[3]{%
    {\vcenter{\hrule height.#3ex\hbox{\vrule width.#2ex height#1ex
     \kern#1ex\vrule width.#3ex}\hrule height.#2ex}}}

\catcode`\@=11 
\newcommand{\parenbar}{\mathpalette\p@renb@r}
\def\p@renb@r#1#2{\vbox{%
  \ifx#1\scriptscriptstyle \dimen@.7em\dimen@ii.2em\else
  \ifx#1\scriptstyle \dimen@.8em\dimen@ii.25em\else
  \dimen@1em\dimen@ii.4em\fi\fi \offinterlineskip
  \ialign{\hfill##\hfill\cr
    \vbox{\hrule width\dimen@ii}\cr
    \noalign{\vskip-.3ex}%
    \hbox to\dimen@{$\mathchar300\hfil\mathchar301$}\cr
    \noalign{\vskip-.3ex}%
    $#1#2$\cr}}}
\catcode`\@=12 

\ifzeusunit
  
\else
  
\fi

\newcommand{\IP}{{\rm I$\kern-0.01667em$P}\xspace}

\mathchardef\qsm=63
\mathchardef\pls=43
\mathchardef\mns=512
\mathchardef\plm=518
\mathchardef\eql=61
\mathchardef\smallleft=300
\mathchardef\smallright=301
\mathchardef\les=316
\mathchardef\gre=318
\mathchardef\leq=532
\mathchardef\grq=533

\catcode`\@=11 
\newcounter{pict@width}
\newcounter{pict@height}
\newlength{\pict@scale}
\newcommand{\psfigadd}[4]{%
\setcounter{pict@width}{1*\ratio{#2+\pict@scale/2}{\pict@scale}}
\setcounter{pict@height}{1*\ratio{#3+\pict@scale/2}{\pict@scale}}
\setlength{\unitlength}{\pict@scale}
\hbox to #2{\hspace{-\fill}\begin{picture}(\thepict@width,\thepict@height)
\put(0,0){\psfig{figure=#1,width=#2,height=#3,clip=}}
\SetScale{0.283466457}
\SetWidth{1.763889}
{#4}
\end{picture}}
}
\newcounter{pict@widthfst}
\newcounter{pict@widthscd}
\newcounter{pict@widthtot}
\newcommand{\psfigaddtwo}[7]{%
\setcounter{pict@widthfst}{1*\ratio{#2+\pict@scale/2}{\pict@scale}}
\setcounter{pict@widthscd}{1*\ratio{#2+#4+\pict@scale/2}{\pict@scale}}
\setcounter{pict@widthtot}{1*\ratio{#2+#4+#6+\pict@scale/2}{\pict@scale}}
\setcounter{pict@height}{1*\ratio{#3+\pict@scale/2}{\pict@scale}}
\setlength{\unitlength}{\pict@scale}
\hbox{\hspace{-\fill}\begin{picture}(\thepict@widthtot,\thepict@height)
\put(0,0){\psfig{figure=#1,width=#2,height=#3,clip=}}
\put(\thepict@widthscd,0){\psfig{figure=#5,width=#6,height=#3,clip=}}
\SetScale{0.283466457}
\SetWidth{1.763889}
{#7}
\end{picture}}
}
\newcommand{\psfigror}[4]{%
\setcounter{pict@width}{1*\ratio{#2+\pict@scale/2}{\pict@scale}}
\setcounter{pict@height}{1*\ratio{#3+\pict@scale/2}{\pict@scale}}
\setlength{\unitlength}{\pict@scale}
\hbox{\begin{picture}(\thepict@width,\thepict@height)
\put(0,\thepict@height){\psfig{figure=#1,width=#3,height=#2,clip=,angle=270}}
\SetScale{0.283466457}
\SetWidth{1.763889}
{#4}
\end{picture}}
}
\newcommand{\psfigrol}[4]{%
\setcounter{pict@width}{1*\ratio{#2+\pict@scale/2}{\pict@scale}}
\setcounter{pict@height}{1*\ratio{#3+\pict@scale/2}{\pict@scale}}
\setlength{\unitlength}{\pict@scale}
\hbox{\begin{picture}(\thepict@width,\thepict@height)
\put(0,0){\psfig{figure=#1,width=#3,height=#2,clip=,angle=90}}
\SetScale{0.283466457}
\SetWidth{1.763889}
{#4}
\end{picture}}
}
\catcode`\@=12 
\newlength\listtextwidth

\catcode`\@=11 
\newlength{\@tabfninsert}
\newlength{\@tabfnwidth}
\newcommand{\tabfootnote}[2]{%
  \setlength{\@tabfninsert}{0.8em}
  \setlength{\@tabfnwidth}{\textwidth}
  \addtolength{\@tabfnwidth}{-\@tabfninsert}
  \addtolength{\@tabfnwidth}{-0.4em}
  \noindent\makebox[\@tabfninsert][r]{\footnotesize$^{#1}$\hfil}\hfill%
  \parbox[t]{\@tabfnwidth}{\footnotesize #2\hfill}}
\catcode`\@=12 

\def\citeCTD{{\cite{%
nim:a279:290,*npps:b32:181,*nim:a338:254%
}}\xspace}
\def\citeMVD{{\cite{%
nim:a581:656%
}}\xspace}
\def\citeSTT{{\cite{%
nim:a535:191%
}}\xspace}
\def\citeCAL{{\cite{%
nim:a309:77,*nim:a309:101,*nim:a321:356,*nim:a336:23%
}}\xspace}
\def\citeHES{{\cite{%
nim:a277:176%
}}\xspace}
\def\citeSixmT{{\cite{%
thesis:gosau:2007%
}}\xspace}
\def\citeBAC{{\cite{%
nim:a300:480%
}}\xspace}
\def\citePCAL{{\cite{%
desy-92-066,*zfp:c63:391,*acpp:b32:2025%
}}\xspace}
\def\citeSPECTRO{{\cite{%
nim:a565:572%
}}\xspace}

\begin{document}

\zeustitle{%
Search for effective Lorentz and CPT
violation using ZEUS data
}

\zeusauthor{ZEUS Collaboration}
\zeusdate{}
\prepnum{DESY--22--107}
\prepdate{December 2022}

\maketitle

%

\begin{abstract}\noindent
  Lorentz and CPT symmetry
  in the quark sector of the Standard Model
  are studied in the context of an effective field
  theory using ZEUS $e^{\pm} p$ data.
  Symmetry-violating effects 
  can lead to time-dependent oscillations of 
  otherwise time-independent observables,
  including scattering cross sections.
  An analysis using five years 
  of inclusive neutral-current
  deep inelastic scattering events
  corresponding to an integrated HERA luminosity
  of $\unit[372]{\pbi}$ at $\sqrt{s} = \unit[318]{\Gev}$
  has been performed. No evidence for 
  oscillations in sidereal time has been 
  observed within statistical and systematic uncertainties. 
  Constraints, most for the first time,
  are placed on 42 coefficients 
  parameterising dominant 
  CPT-even dimension-four 
  and CPT-odd dimension-five
  spin-independent modifications
  to the propagation and interaction 
  of light quarks. 

\end{abstract}

\thispagestyle{empty}
\cleardoublepage
%
%
%
%


\topmargin-1.cm
\evensidemargin-0.3cm
\oddsidemargin-0.3cm
\textwidth 16.cm
\textheight 680pt
\parindent0.cm
\parskip0.3cm plus0.05cm minus0.05cm
\def\3{\ss}
\newcommand{\address}{ }
\pagenumbering{Roman}
                                                   %
\begin{center}
{                      \Large  The ZEUS Collaboration              }
\end{center}

{\small\raggedright


I.~Abt$^{1}$, 
R. Aggarwal$^{2}$, 
V.~Aushev$^{3}$, 
O.~Behnke$^{4}$, 
A.~Bertolin$^{5}$, 
I.~Bloch$^{6}$, 
I.~Brock$^{7}$, 
N.H.~Brook$^{8, a}$, 
R.~Brugnera$^{9}$, 
A.~Bruni$^{10}$, 
P.J.~Bussey$^{11}$, 
A.~Caldwell$^{1}$, 
C.D.~Catterall$^{12}$, 
J.~Chwastowski$^{13}$, 
J.~Ciborowski$^{14, b}$, 
R.~Ciesielski$^{4, c}$, 
A.M.~Cooper-Sarkar$^{15}$, 
M.~Corradi$^{10, d}$, 
R.K.~Dementiev$^{16}$, 
S.~Dusini$^{5}$, 
J.~Ferrando$^{4}$, 
B.~Foster$^{15, e}$, 
E.~Gallo$^{17, f}$, 
D.~Gangadharan$^{18, g}$, 
A.~Garfagnini$^{9}$, 
A.~Geiser$^{4}$, 
G.~Grzelak$^{14}$, 
C.~Gwenlan$^{15}$, 
D.~Hochman$^{19}$, 
N.Z.~Jomhari$^{4}$, 
I.~Kadenko$^{3}$, 
U.~Karshon$^{19}$, 
P.~Kaur$^{20}$, 
R.~Klanner$^{17}$, 
I.A.~Korzhavina$^{16}$, 
N.~Kovalchuk$^{17}$, 
M.~Kuze$^{21}$, 
B.B.~Levchenko$^{16}$, 
A.~Levy$^{22}$, 
B.~L\"ohr$^{4}$, 
E.~Lohrmann$^{17}$, 
A.~Longhin$^{9}$, 
F.~Lorkowski$^{4}$,
E.~Lunghi$^{23}$,
I.~Makarenko$^{4}$, 
J.~Malka$^{4, h}$, 
S.~Masciocchi$^{24, i}$, 
K.~Nagano$^{25}$, 
J.D.~Nam$^{26}$, 
Yu.~Onishchuk$^{3}$, 
E.~Paul$^{7}$, 
I.~Pidhurskyi$^{27}$, 
A.~Polini$^{10}$, 
M.~Przybycie\'n$^{28}$, 
A.~Quintero$^{25}$, 
M.~Ruspa$^{29}$, 
U.~Schneekloth$^{4}$, 
T.~Sch\"orner-Sadenius$^{4}$, 
I.~Selyuzhenkov$^{24}$, 
M.~Shchedrolosiev$^{4}$, 
L.M.~Shcheglova$^{16}$,
N.~Sherrill$^{30}$,
I.O.~Skillicorn$^{11}$, 
W.~S{\l}omi\'nski$^{31}$, 
A.~Solano$^{32}$, 
L.~Stanco$^{5}$, 
N.~Stefaniuk$^{4}$, 
B.~Surrow$^{26}$, 
K.~Tokushuku$^{25}$, 
O.~Turkot$^{4, h}$, 
T.~Tymieniecka$^{33}$, 
A.~Verbytskyi$^{1}$, 
W.A.T.~Wan Abdullah$^{34}$, 
K.~Wichmann$^{4}$, 
M.~Wing$^{8, j}$, 
S.~Yamada$^{25}$, 
Y.~Yamazaki$^{35}$, 
A.F.~\.Zarnecki$^{14}$, 
O.~Zenaiev$^{4, k}$ 
\newpage


{\setlength{\parskip}{0.4em}
\makebox[3ex]{$^{1}$}
\begin{minipage}[t]{14cm}
{\it Max-Planck-Institut f\"ur Physik, M\"unchen, Germany}

\end{minipage}

\makebox[3ex]{$^{2}$}
\begin{minipage}[t]{14cm}
{\it DST-Inspire Faculty, Department of Technology, SPPU, India}

\end{minipage}

\makebox[3ex]{$^{3}$}
\begin{minipage}[t]{14cm}
{\it Department of Nuclear Physics, National Taras Shevchenko University of Kyiv, Kyiv, Ukraine}

\end{minipage}

\makebox[3ex]{$^{4}$}
\begin{minipage}[t]{14cm}
{\it Deutsches Elektronen-Synchrotron DESY, Notkestr. 85, 22607 Hamburg, Germany}

\end{minipage}

\makebox[3ex]{$^{5}$}
\begin{minipage}[t]{14cm}
{\it INFN Padova, Padova, Italy}~$^{A}$

\end{minipage}

\makebox[3ex]{$^{6}$}
\begin{minipage}[t]{14cm}
{\it Deutsches Elektronen-Synchrotron DESY, Platanenallee 6, 15738 Zeuthen, Germany}
\end{minipage}

\makebox[3ex]{$^{7}$}
\begin{minipage}[t]{14cm}
{\it Physikalisches Institut der Universit\"at Bonn,
Bonn, Germany}~$^{B}$

\end{minipage}

\makebox[3ex]{$^{8}$}
\begin{minipage}[t]{14cm}
{\it Physics and Astronomy Department, University College London,
London, United Kingdom}~$^{C}$

\end{minipage}

\makebox[3ex]{$^{9}$}
\begin{minipage}[t]{14cm}
{\it Dipartimento di Fisica e Astronomia dell' Universit\`a and INFN,
Padova, Italy}~$^{A}$

\end{minipage}

\makebox[3ex]{$^{10}$}
\begin{minipage}[t]{14cm}
{\it INFN Bologna, Bologna, Italy}~$^{A}$

\end{minipage}

\makebox[3ex]{$^{11}$}
\begin{minipage}[t]{14cm}
{\it School of Physics and Astronomy, University of Glasgow,
Glasgow, United Kingdom}~$^{C}$

\end{minipage}

\makebox[3ex]{$^{12}$}
\begin{minipage}[t]{14cm}
{\it Department of Physics, York University, Ontario, Canada M3J 1P3}~$^{D}$

\end{minipage}

\makebox[3ex]{$^{13}$}
\begin{minipage}[t]{14cm}
{\it The Henryk Niewodniczanski Institute of Nuclear Physics, Polish Academy of \\
Sciences, Krakow, Poland}

\end{minipage}

\makebox[3ex]{$^{14}$}
\begin{minipage}[t]{14cm}
{\it Faculty of Physics, University of Warsaw, Warsaw, Poland}

\end{minipage}

\makebox[3ex]{$^{15}$}
\begin{minipage}[t]{14cm}
{\it Department of Physics, University of Oxford,
Oxford, United Kingdom}~$^{C}$

\end{minipage}

\makebox[3ex]{$^{16}$}
\begin{minipage}[t]{14cm}
{\it Affiliated with an institute covered by a current or former 
collaboration agreement with DESY}

\end{minipage}

\makebox[3ex]{$^{17}$}
\begin{minipage}[t]{14cm}
{\it Hamburg University, Institute of Experimental Physics, Hamburg,
Germany}~$^{E}$

\end{minipage}

\makebox[3ex]{$^{18}$}
\begin{minipage}[t]{14cm}
{\it Physikalisches Institut of the University of Heidelberg, Heidelberg, Germany}

\end{minipage}

\makebox[3ex]{$^{19}$}
\begin{minipage}[t]{14cm}
{\it Department of Particle Physics and Astrophysics, Weizmann
Institute, Rehovot, Israel}

\end{minipage}

\makebox[3ex]{$^{20}$}
\begin{minipage}[t]{14cm}
{\it Sant Longowal Institute of Engineering and Technology, Longowal, Punjab, India}

\end{minipage}

\makebox[3ex]{$^{21}$}
\begin{minipage}[t]{14cm}
{\it Department of Physics, Tokyo Institute of Technology,
Tokyo, Japan}~$^{F}$

\end{minipage}

\makebox[3ex]{$^{22}$}
\begin{minipage}[t]{14cm}
{\it Raymond and Beverly Sackler Faculty of Exact Sciences, School of Physics, \\
Tel Aviv University, Tel Aviv, Israel}~$^{G}$

\end{minipage}

\makebox[3ex]{$^{23}$}
\begin{minipage}[t]{14cm}
{\it   Department of Physics, Indiana University Bloomington, Bloomington, IN 47405, USA}

\end{minipage}

\makebox[3ex]{$^{24}$}
\begin{minipage}[t]{14cm}
{\it GSI Helmholtzzentrum f\"{u}r Schwerionenforschung GmbH, Darmstadt, Germany}

\end{minipage}

\makebox[3ex]{$^{25}$}
\begin{minipage}[t]{14cm}
{\it Institute of Particle and Nuclear Studies, KEK,
Tsukuba, Japan}~$^{F}$

\end{minipage}

\makebox[3ex]{$^{26}$}
\begin{minipage}[t]{14cm}
{\it Department of Physics, Temple University, Philadelphia, PA 19122, USA}~$^{H}$

\end{minipage}

\makebox[3ex]{$^{27}$}
\begin{minipage}[t]{14cm}
{\it Institut f\"ur Kernphysik, Goethe Universit\"at, Frankfurt am Main, Germany}

\end{minipage}

\makebox[3ex]{$^{28}$}
\begin{minipage}[t]{14cm}
{\it AGH University of Science and Technology, Faculty of Physics and Applied Computer
Science, Krakow, Poland}

\end{minipage}

\makebox[3ex]{$^{29}$}
\begin{minipage}[t]{14cm}
{\it Universit\`a del Piemonte Orientale, Novara, and INFN, Torino,
Italy}~$^{A}$

\end{minipage}

\makebox[3ex]{$^{30}$}
\begin{minipage}[t]{14cm}
{\it Department of Physics and Astronomy, 
University of Sussex, Brighton, BN1 9QH, United Kingdom}~$^{I}$

\end{minipage}

\makebox[3ex]{$^{31}$}
\begin{minipage}[t]{14cm}
{\it Department of Physics, Jagellonian University, Krakow, Poland}~$^{J}$

\end{minipage}

\makebox[3ex]{$^{32}$}
\begin{minipage}[t]{14cm}
{\it Universit\`a di Torino and INFN, Torino, Italy}~$^{A}$

\end{minipage}

\makebox[3ex]{$^{33}$}
\begin{minipage}[t]{14cm}
{\it National Centre for Nuclear Research, Warsaw, Poland}

\end{minipage}

\makebox[3ex]{$^{34}$}
\begin{minipage}[t]{14cm}
{\it National Centre for Particle Physics, Universiti Malaya, 50603 Kuala Lumpur, Malaysia}~$^{K}$

\end{minipage}

\makebox[3ex]{$^{35}$}
\begin{minipage}[t]{14cm}
{\it Department of Physics, Kobe University, Kobe, Japan}~$^{F}$

\end{minipage}

}

\vspace{3em}


{\setlength{\parskip}{0.4em}\raggedright
\makebox[3ex]{$^{ A}$}
\begin{minipage}[t]{14cm}
 supported by the Italian National Institute for Nuclear Physics (INFN) \
\end{minipage}

\makebox[3ex]{$^{ B}$}
\begin{minipage}[t]{14cm}
 supported by the German Federal Ministry for Education and Research (BMBF), under
 contract No.\ 05 H09PDF\
\end{minipage}

\makebox[3ex]{$^{ C}$}
\begin{minipage}[t]{14cm}
 supported by the Science and Technology Facilities Council, UK\
\end{minipage}

\makebox[3ex]{$^{ D}$}
\begin{minipage}[t]{14cm}
 supported by the Natural Sciences and Engineering Research Council of Canada (NSERC) \
\end{minipage}

\makebox[3ex]{$^{ E}$}
\begin{minipage}[t]{14cm}
 supported by the German Federal Ministry for Education and Research (BMBF), under
 contract No.\ 05h09GUF, and the SFB 676 of the Deutsche Forschungsgemeinschaft (DFG) \
\end{minipage}

\makebox[3ex]{$^{ F}$}
\begin{minipage}[t]{14cm}
 supported by the Japanese Ministry of Education, Culture, Sports, Science and Technology
 (MEXT) and its grants for Scientific Research\
\end{minipage}

\makebox[3ex]{$^{ G}$}
\begin{minipage}[t]{14cm}
 supported by the Israel Science Foundation\
\end{minipage}

\makebox[3ex]{$^{ H}$}
\begin{minipage}[t]{14cm}
 supported in part by the Office of Nuclear Physics within the U.S.\ DOE Office of Science
\end{minipage}

\makebox[3ex]{$^{ I}$}
\begin{minipage}[t]{14cm}
supported in part by the Science and Technology Facilities Council grant number ST/T006048/1
\end{minipage}

\makebox[3ex]{$^{ J}$}
\begin{minipage}[t]{14cm}
supported by the Polish National Science Centre (NCN) grant no.\ DEC-2014/13/B/ST2/02486
\end{minipage}

\makebox[3ex]{$^{ K}$}
\begin{minipage}[t]{14cm}
 supported by HIR grant UM.C/625/1/HIR/149 and UMRG grants RU006-2013, RP012A-13AFR and RP012B-13AFR from
 Universiti Malaya, and ERGS grant ER004-2012A from the Ministry of Education, Malaysia\
\end{minipage}

}

\pagebreak[4]
{\setlength{\parskip}{0.4em}


\makebox[3ex]{$^{ a}$}
\begin{minipage}[t]{14cm}
now at University of Bath, United Kingdom\
\end{minipage}

\makebox[3ex]{$^{ b}$}
\begin{minipage}[t]{14cm}
also at Lodz University, Poland\
\end{minipage}

\makebox[3ex]{$^{ c}$}
\begin{minipage}[t]{14cm}
now at Rockefeller University, New York, NY 10065, USA\
\end{minipage}

\makebox[3ex]{$^{ d}$}
\begin{minipage}[t]{14cm}
now at INFN Roma, Italy\
\end{minipage}

\makebox[3ex]{$^{ e}$}
\begin{minipage}[t]{14cm}
also at DESY and University of Hamburg, Hamburg, Germany and supported by a Leverhulme Trust Emeritus Fellowship\
\end{minipage}

\makebox[3ex]{$^{ f}$}
\begin{minipage}[t]{14cm}
also at DESY, Hamburg, Germany\
\end{minipage}

\makebox[3ex]{$^{ g}$}
\begin{minipage}[t]{14cm}
now at University of Houston, Houston, TX 77004, USA\
\end{minipage}


\makebox[3ex]{$^{ h}$}
\begin{minipage}[t]{14cm}
now at European X-ray Free-Electron Laser facility GmbH, Hamburg, Germany\
\end{minipage}

\makebox[3ex]{$^{ i}$}
\begin{minipage}[t]{14cm}
also at Physikalisches Institut of the University of Heidelberg, Heidelberg,  Germany\
\end{minipage}

\makebox[3ex]{$^{ j}$}
\begin{minipage}[t]{14cm}
also supported by DESY, Hamburg, Germany\
\end{minipage}

\makebox[3ex]{$^{ k}$}
\begin{minipage}[t]{14cm}
now at Hamburg University, II. Institute for Theoretical Physics, Hamburg, Germany \
\end{minipage}


}

}


 \cleardoublepage
\pagenumbering{arabic}
%
%
\section{Introduction}
\label{sec:intro}
Relativity is one of the best
established principles in physics. 
It concerns the invariance
of physical laws under 
transformations of spacetime orientation
including spatial 
rotations and velocity boosts.
As a core principle
of classical and modern theories,
relativity implies that
identical measurements
performed with different
spacetime orientations
observe the same
laws of motion, with 
their respective results linked
by the appropriate transformation.
In general, any theory 
exhibiting isotropy
under rotations and relativistic boosts 
is said to be Lorentz invariant. 

Searches for violations of 
isotropy date back to the 
Michelson--Morley experiment,
which attempted to measure 
the rotational anisotropy 
of light propagation~\cite{Michelson:1887zz}.
Its null result 
heavily influenced 
the support of
special relativity.
As rotations and boosts
do not commute, 
the violation of 
rotation invariance implies
the violation of boost invariance
and vice versa.
However, experimental 
indications of such symmetry violations
do not necessarily
imply Lorentz violation 
in an underlying fundamental theory.
This can be seen, for example, 
by considering the Earth's motion 
in the presence of a hypothetical 
Lorentz-invariant background field, 
perhaps representing
a galactic dark-matter halo. 
If the field's velocity distribution is isotropic
in the galactic frame as is commonly assumed,
it will be anisotropic
in an Earth-based laboratory frame, 
leading to locally time-dependent observations
primarily as a function of the Earth's 
annual revolution around the Sun and its axial rotation.
Lorentz violation implies CPT violation
in many theoretical frameworks.
Generic searches for violations
of rotation invariance 
encompass a wide range of possible physical effects.

Model-independent experimental tests 
of Lorentz invariance and related
fundamental symmetries 
have been intensively performed 
for over two decades, with no 
significant deviations observed~\cite{tables}.
In spite of the substantial progress achieved in 
understanding possible Lorentz and CPT violation,
relatively few 
studies involving the quark sector 
directly have been performed, 
leaving unexamined a 
vast array of potential signals.
One reason for this 
stems from the difficulty in interpreting
quark-level interactions in stable-hadron 
and lepton processes.
A few previous collider-based searches
have resulted in constraints 
on renormalisable 
and rotationally invariant 
quark-sector effects in the final states of
$e^+e^-$ collisions using LEP data~\cite{Karpikov:2016qvq}
and on top-quark effects~\cite{Berger:2015yha}
using Tevatron data~\cite{D0:2012rbu}.
Studies of Lorentz-violating effects in 
hadronic processes have recently 
been addressed in a series
of theoretical studies on DIS~\cite{klv17,ls18,kl19} 
and the Drell--Yan process
~\cite{klsv19,lssv21}. 

This paper describes
a search for effective
rotation-violating signals 
affecting light quarks ($f = u,d,s$) 
performed with
$e^\pm p$ data collected 
by the ZEUS detector at HERA. 
Deep inelastic scattering (DIS) is chosen 
as a suitable test process 
because of its strong theoretical
foundation and the wealth of available data.
To remain model independent, 
the technical framework employed 
is rooted in an
effective field theory (EFT)~\cite{sw}
and is reviewed in the following section.
Modified DIS cross sections and related
observables incorporating the 
effects of interest are then described,
followed by the experimental
setup, analysis method and 
associated Monte Carlo studies, 
and a description of 
systematic uncertainties. 
Constraints on effective couplings
parameterising signals that violate rotational symmetry
in light-quark
interactions are given.

\section{Theory}
\label{sec:theory}

\subsection{The Standard Model Extension}
\label{ssec:SME}
Searches for Lorentz-violating effects often 
make use of 
the comprehensive EFT framework known as 
the Standard Model Extension (SME)~\cite{ck97,
ck98,ak04,Kostelecky:2020hbb}.
In the absence of general-relativistic effects,
the SME action may be written as
\begin{equation}
\label{SMEaction}
S_{\text{SME}} = S_{\text{SM}} + S_{\text{LV}},
\end{equation}
where $S_{\text{SM}}$ is the action 
of the Standard Model
and $S_{\text{LV}}$ represents all possible 
Lorentz-violating terms constructed
from additional interactions 
of Standard Model fields.
These terms are typically treated as 
perturbations with respect to conventional effects.
The analysis presented here is
restricted to selected dimension-four
and -five operators.
As CPT violation implies Lorentz violation 
in a unitary and local quantum field 
theory~\cite{ck97,owg}, 
CPT-violating
operators are also contained in the SME.
Further information on the SME can be found in, e.g., 
accessible reviews~\cite{review1,review2}.

The dimension-four operators,
$\bar{\psi}_{f} 
\gamma_\mu iD_{\nu}\psi_{f}$,
considered here are the dominant
spin-independent 
and renormalisable quark-sector 
modifications of quantum electrodynamics~\cite{klv17}
\footnote{The notation h.c. is an abbreviation for 
Hermitian conjugation.},
\begin{equation}
\mathcal{L}_c = 
\tfrac{1}{2}\sum_{f}c_f^{\mu\nu}\bar{\psi}_{f} 
\gamma_\mu iD_{\nu}\psi_{f} + \text{h.c.}\;,
\label{eq:cmodel}
\end{equation}
where $D_\nu = \partial_\nu + i e e_{\hspace{-0.3mm}f} A_\nu$, 
$e$ is the electron charge, $e_{\hspace{-0.3mm}f}$ are 
the relative quark charges and
$A_\nu$ is the photon field.
The coefficients $c_f^{\mu\nu}$ control 
the magnitude of Lorentz-violating effects.
Since the coefficient and operator 
Lorentz indices in Eq.~\eqref{eq:cmodel}
are contracted, $\mathcal{L}_c$ is 
invariant under coordinate transformations, 
known as "observer transformations"~\cite{ck97}. 
Accordingly, observer transformations 
have no sensitivity to anisotropies
parameterised by $c_f^{\mu\nu}$.
If instead a "particle transformation" is performed, 
the particle fields $\psi_f, A_\nu, \ldots$ are
transformed while the coordinates are unaffected. 
This could be realised,
for example, by physically rotating the system described
by the particle fields.
In this case,
$\mathcal{L}_{\text{c}}$ is not
Lorentz invariant in general.
This occurs because $c_f^{\mu\nu}$
is invariant under particle transformations, whereas the 
system described by the particle fields 
is not.
A distinction between
transforming the system as opposed 
to transforming the observer is present. 
This violates Lorentz invariance.
These $c$-type coefficients may be 
taken as symmetric and traceless, 
leaving nine observable coefficients per flavour\;$f$~\cite{ck98}. 

All gauge-invariant effects 
in the minimal and non-minimal
sector of non-Abelian gauge theories 
were classified and dimension-five, 
spin-independent, and CPT-violating 
effects 
\begin{equation}
\mathcal{L}_{a^{(5)}} =
-\tfrac{1}{2}\sum_{f}{a^{(5)}_{f}}^{\mu\alpha\beta}
\bar{\psi}_{f}\gamma_\mu 
i D_{(\alpha}i  D_{\beta)}\psi_{f} + \text{h.c.}\;,
\label{eq:a5model}
\end{equation}
parameterised
by the $a^{(5)}$-type coefficients 
$a_f^{(5)\mu\alpha\beta}$,
were considered for DIS~\cite{kl19}. Note 
that $\mathcal{L}_{a^{(5)}}$ represents
the dominant non-minimal and 
spin-independent CPT-violating effects
on quarks, and the parenthesis 
notation denotes symmetrisation
with respect to the indices $\alpha, \beta$. 
Similar to the case of the 
$c$-type coefficients, 
the $a^{(5)}$-type coefficients
may be taken to be totally symmetric and 
traceless~\cite{Edwards:2019lfb,Ding:2016lwt,klsv19}.
This subset of coefficients is denoted 
$a_{\text{S}f}^{(5)\mu\alpha\beta}$ and contains
16 independent observable components per flavour.

Simulations have been performed 
assessing the sensitivity of the $c$-type
and $a^{(5)}$-type coefficients 
for quarks and antiquarks
as measured in DIS data from HERA~\cite{hera}, 
Electron--Ion Collider pseudodata~\cite{ls18}, 
and in the Drell--Yan process using
LHC data~\cite{Sirunyan:2018owv}.
\subsection{Adding SME effects to the description of DIS}
\label{ssec:DIS}
The class of operators that modify the free propagation 
of quarks and their covariant couplings
to gauge fields was examined~\cite{klsv19}, 
resulting in the development of a tree-level,
Lorentz-violating version of the parton model. 
This leads to a modified description of DIS.

In the following, the external 
momenta of the incident lepton and
proton are denoted by 
$l^\mu$ and $p^\mu$, respectively, while 
the scattered lepton and exchanged momentum 
are denoted by $l'^\mu$ and $q^\mu = l^\mu - l'^\mu$, 
where $-q^2 \equiv Q^2$ is the momentum transfer. 
The Bjorken variables are
\begin{equation}
x_{\text{Bj}} = \frac{Q^2}{2p\cdot q}\;, 
\quad y_{\text{Bj}} = \frac{p\cdot q}{p\cdot l}\;,
\end{equation}
where $s\approx 2p\cdot l$ and 
$Q^2 \approx x_{\text{Bj}}y_{\text{Bj}}s$. In the DIS limit,
and working at zeroth order in the strong coupling constant,
the inclusion of Lorentz-violating effects 
described by Eq.~\eqref{eq:cmodel} at tree level 
results in the cross section\footnote{Equation~\eqref{xsecc} reduces
to the leading-order Standard Model cross section
in the limit $c_f^{\mu\nu}\rightarrow 0$.}
\begin{equation}
\frac{d\sigma}{dx_{\text{Bj}} dy_{\text{Bj}} d\varphi} = 
\frac{\alpha^2y_{\text{Bj}}}{2 Q^4}\sum_{f} e_{\hspace{-0.3mm}f}^2\frac{1}{\widetilde{Q}_f^2} 
L_{\mu\nu}H_f^{\mu\nu}f_f(\widetilde{x}_f)\;,
\label{xsecc}
\end{equation}
where $\varphi$ is the scattered lepton azimuthal angle, 
$\widetilde{Q}^2_f = -\widetilde{q}_f^2 = -(q^\mu + c_f^{\mu q})(q_\mu + c_{f\mu q})$
where $c_f^{\mu q} \equiv c_f^{\mu\nu}q_\nu$, and 
\begin{eqnarray}
L_{\mu\nu}H_f^{\mu\nu} &=& 
8 \left[2(\widehat{k}_f\cdot l)(\widehat{k}_f\cdot l') 
+ \widehat{k}_f\cdot(l-l')(l\cdot l') 
+ 2(\widehat{k}_f\cdot l)\left(c_f^{\widehat{k}_fl'} 
+ c_f^{l'\widehat{k}_f} - c_f^{l'l'} \right) 
\right.
\nonumber\\
& &
\hskip 0.5cm \left. 
+ 2(\widehat{k}_f\cdot l')\left(c_f^{\widehat{k}_fl} 
+ c_f^{l\widehat{k}_f} + c_f^{ll}\right) 
- 2(l\cdot l')c_f^{\widehat{k}_f\widehat{k}_f}\right],
\label{eq:Hfc}
\end{eqnarray}
with $\widehat{k}_f^\mu = \widetilde{x}_f(p^\mu-c_f^{\mu p})$. 
The parton distribution functions (PDFs) 
are denoted $f_f(\widetilde{x}_f)$ and are 
evaluated at the shifted Bjorken variable~\cite{klsv19}
\begin{equation}
\widetilde{x}_f = x_{\text{Bj}}\left(1 + \frac{2c_f^{qq}}{q^2} \right) 
+ \frac{x_{\text{Bj}}^2}{q^2}\left(c_f^{pq} + c_f^{qp}\right),
\label{xprimec}
\end{equation}
where $c_f^{qq} \equiv c_f^{\mu\nu}q_{\mu}q_{\nu}$, etc. 
The PDFs are taken to be
the conventional leading-order 
Standard Model PDFs evaluated at $\widetilde{x}_f$ 
using the MSTW 2008 variable-flavour PDF set~\cite{Martin:2009iq},
as implemented in the program 
ManeParse~\cite{Maneparse1,maneparse}.\footnote{
It was checked that effects of a variation of the input PDF set
are negligible. Potential effects from Lorentz violation on the PDF starting scale parametrization and on the PDF evolution were neglected.
}
Note that since the $c$-type 
coefficients control CPT-even effects, 
the quark and antiquark contributions are identical 
in the cross section Eq.~\eqref{xsecc}, except
for the PDF dependence. Thus, for
these coefficients, 
$d\sigma \propto (f_f + f_{\bar f})$ and the dominant sensitivity
is expected at low $x_{\text{Bj}}$.

\begin{figure}[p]
\begin{center}
\includegraphics[width=0.6 \linewidth]
{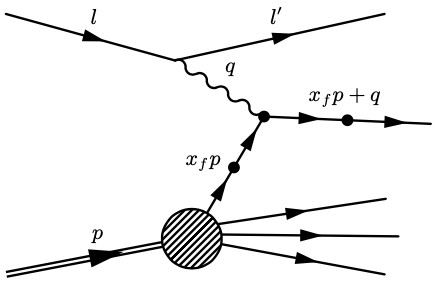}
\end{center}
\caption{Parton-model picture of deep inelastic 
scattering with Lorentz violation. The dots on 
the incident parton, struck parton, and at 
the photon-parton vertex correspond to 
modified propagation and interaction due 
to Lorentz violation. The parton 
carries a momentum fraction $x_f$
that is perturbed from $x_{\textnormal{Bj}}$
by the coefficients
for Lorentz violation at first order.}
\label{fig:partonscattering}
\end{figure}

The cross section including effects
of the $a^{(5)}$-type
coefficients from Eq.~\eqref{eq:a5model} is~\cite{klv17,klsv19}
\begin{eqnarray}
\frac{d\sigma}{dx_{\text{Bj}}dy_{\text{Bj}} d\varphi} 
&=& \frac{\alpha^2}{Q^4} \sum_{f} F_{2f}
\left[\frac{y_{\text{Bj}}s^2}{\pi}
\left[1+(1-y_{\text{Bj}})^2\right]\delta_{\text{S}f} 
+ \frac{y_{\text{Bj}}(y_{\text{Bj}}-2)s}{x_{\text{Bj}}}x_{\text{S}f} 
\right. 
\nonumber\\
& & \left. 
\hskip 30pt
-\frac{4}{x_{\text{Bj}}}
\left(4x_{\text{Bj}}^2a_{\text{S}f}^{(5)ppl} 
+ 6x_{\text{Bj}}a_{\text{S}f}^{(5)lpq} 
+ 2a_{\text{S}f}^{(5)lqq}\right) 
\right. 
\nonumber\\
&&  \left. 
\hskip 30pt
+2y_{\text{Bj}}\left(4x_{\text{Bj}}^2a_{\text{S}f}^{(5)ppp} 
+ 4x_{\text{Bj}}a_{\text{S}f}^{(5)ppq} 
+ 4x_{\text{Bj}}a_{\text{S}f}^{(5)lpp} 
+ 2a_{\text{S}f}^{(5)lpq} 
+ a_{\text{S}f}^{(5)pqq}\right) 
\right. 
\nonumber\\
&&\left. 
\hskip 30pt
+ \frac{4y_{\text{Bj}}}{x_{\text{Bj}}}
\left(2x_{\text{Bj}}a_{\text{S}f}^{(5)llp} 
+ a_{\text{S}f}^{(5)llq}\right)\right],
\label{xseca5}
\end{eqnarray}
where $F_{2f} = e_{\hspace{-0.3mm}f}^2 f_f(x_{\text{S}f}')x_{\text{S}f}'$ 
with $x_{\text{S}f}' = x_{\text{Bj}} - x_{\text{S}f}$ and 
\begin{eqnarray}
\delta_{\text{S}f} &=& 
\frac{\pi}{y_{\text{Bj}}s} 
\left[1+\frac{2}{y_{\text{Bj}}s} 
\left(4x_{\text{Bj}} a_{\text{S}f}^{(5)ppq} 
+ 3a_{\text{S}f}^{(5)pqq} 
\right)\right], \\
x_{\text{S}f} & = &
-\frac{2}{y_{\text{Bj}}s}
\left(2x_{\text{Bj}}^2 a_{\text{S}f}^{(5)ppq} 
+ 3x_{\text{Bj}}a_{\text{S}f}^{(5)pqq} 
+ a_{\text{S}f}^{(5)qqq}\right), \label{xprimea5} \\
a_{{\text S}f}^{(5)\mu\alpha\beta} & =  & 
\tfrac 13 \sum_{(\mu\alpha\beta)} 
( a_{f}^{(5)\mu\alpha\beta} 
- \tfrac 16 a_{f}^{(5)\mu\rho\sigma} \eta_{\rho\sigma} \eta^{\alpha\beta}
- \tfrac 13 a_{f}^{(5)\rho\mu\sigma} \eta_{\rho\sigma} \eta^{\alpha\beta}) \label{eq:a5S},
\end{eqnarray}
where $a_{\text{S}f}^{(5)qqq} \equiv 
a_{\text{S}f}^{(5)\mu\alpha\beta}q_\mu q_\alpha q_\beta$, etc. 
The sum in Eq.~\eqref{eq:a5S} denotes
symmetrisation with respect to the
indices $\mu, \alpha$ and $\beta$,
and $\eta_{\rho\sigma}$ is the Minkowski
metric.
As the operator in Eq.~\eqref{eq:a5model} 
is odd under a CPT transformation, for the
antiparticle contributions to the 
sum in Eq.~\eqref{xseca5}, a replacement 
$a_{\text{S}f}^{(5)\mu\alpha\beta} \rightarrow 
-a_{\text{S}f}^{(5)\mu\alpha\beta}$ for antiquark flavours
must be performed, along with multiplication 
of the appropriate PDFs.
In contrast to the $c$-type coefficients, 
this feature implies $d\sigma \propto (f_f - f_{\bar f})$, thus
giving weaker sensitivity at low $x_{\text{Bj}}$. 
A depiction of the DIS process with Lorentz violation
is shown in Fig.~\ref{fig:partonscattering}.

\subsection{Sidereal signals}
\label{ssec:sidsignals}
An experiment is sensitive
to the $c$- and $a^{(5)}$-type coefficients as 
they appear in the laboratory frame.
The cross sections Eqs.~\eqref{xsecc} and \eqref{xseca5} therefore 
must be evaluated in the laboratory
frame of the ZEUS detector.
This frame is non-inertial due to the Earth's
axial rotation and revolution around the Sun. 
By convention, an approximately inertial 
frame with spatial coordinates fixed on the centre of the Sun, 
known as the Sun-centred frame (SCF),
is introduced as a convenient frame to report 
constraints on the SME coefficients~\cite{km02,bklr02,bklr03}.
In the SCF, the coefficients 
are typically assumed to be 
spacetime constants, implying 
the preservation of translation invariance
and the conservation of four-momentum~\cite{ak04}. 
Thus, in another frame rotating with respect to the SCF, 
such as the laboratory frame,
the coefficients will oscillate as a function of time.
To make this time dependence explicit,
the coefficients appearing in the laboratory-frame 
cross sections are re-expressed in terms of the SCF coefficients by performing
the appropriate Lorentz transformation $\Lambda^{\mu}_{\hphantom{\mu}\nu}$. 
For example, the transformation for 
the $c$-type coefficients is
\begin{equation}
\label{eq:ctransformation}
c_{\text{lab}}^{\mu\nu} = 
\Lambda^\mu_{\hphantom{\mu}\vphantom{0}\alpha}\Lambda^\nu_{\hphantom{\nu}\vphantom{0}\beta}c_{\text{SCF}}^{\alpha\beta}.
\end{equation} 

In what follows, coefficients that appear
with Greek and numeric indices 
denote laboratory-frame
coefficients and those 
with capital Latin indices 
denote SCF coefficients.
After performing this transformation and re-expressing 
the laboratory-frame coefficients, 
the laboratory cross sections 
are functions of the fixed SCF coefficients
and sinusoidal functions with periods
controlled by the length of the sidereal day.
The boost of the Earth 
with respect to the Sun, $\beta_\oplus \approx 10^{-4}$,
is suppressed relative to the 
effect of Earth's rotation by several
orders of magnitude and can be neglected. Also, the 
fact that the Earth's orbit is translated with respect
to the SCF has no physical effect because of translation invariance. 
Therefore, the transformation 
$\Lambda^\mu_{\hphantom{\mu} \vphantom{0}\nu}$
in Eq.~\eqref{eq:ctransformation} is well approximated
as a pure rotation: $\Lambda^0_{\hphantom{0}0} = 1$, 
$\Lambda^\mu_{\hphantom{\mu}0} = 0 =\Lambda^0_{\hphantom{\nu}0}$ and 
$\Lambda^i_{\hphantom{i}j} = 
R^i_{\hphantom{i}j} \; (i,j=1,2,3)$ is an orthogonal matrix.

\begin{figure}[p]
\begin{center}
\includegraphics[width=\linewidth]{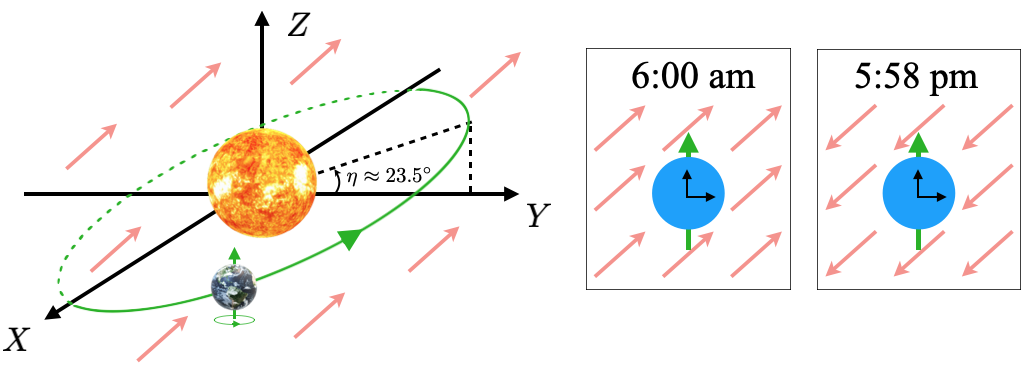}\\
{\scriptsize (a)} \hskip 8.55cm {\scriptsize (b)} \\ \; \\
\end{center}
\caption{Sun-centred frame (SCF) (a). 
The orbit of the Earth in the SCF is shown 
in green. As the $Z$ axis is parallel to 
the Earth's rotation axis, the orbit is 
inclined by the angle $\eta$. The effect of 
Lorentz violation is depicted as red background 
arrows. A laboratory on Earth studying a particle with spin 
(blue disk and green arrow) with 
coordinates given by black arrows centred on 
the particle will observe 
background configurations as 
a function of the sidereal period 
$T_{\textnormal{sidereal}} 
\approx 23$\;\textnormal{h}\;$56$\;\textnormal{min} (b).
The times depicted are an illustration
of the sidereal effect.}
\label{fig:SCF}
\end{figure}

The SCF is depicted in Fig.~\ref{fig:SCF}.
It is defined by 
coordinates $X^\mu = (T,X,Y,Z)$ as follows:
$T=0$ is identified with the date of the 2000 vernal equinox, 
March 20, 2000 at 7:35 UTC; 
the $Z$ axis is aligned with the Earth's rotation axis;
the $X$ axis points from the Earth to the Sun at $T=0$;
and the $Y$ axis completes the right-handed coordinate system.
At $T=0$, the Earth's equator lies in the $XY$ plane
and the longitude $\lambda_0 \approx 66.25^\circ$ 
will observe the Sun directly 
overhead, towards the local zenith.
Therefore, $T=0$ 
is a suitable moment for easily 
relating the SCF coordinates to the 
laboratory-frame coordinates.
The small effects of the 
Earth's non-circular orbit can be neglected.
For the ZEUS detector, 
the co-latitude, orientation of the electron/positron-
or proton-beam direction 
and the local sidereal time $T_\oplus$ must be specified. 
The zero of $T_\oplus$ is defined
as one of the moments when 
the $y$ axis of the laboratory
is parallel to the SCF $Y$ axis, which on 
the date of the equinox occurs
shortly after $T=0$. 
In other words, since HERA is not situated at the longitude $\lambda_0$, 
but the longitude 
$\lambda \approx 9.88^\circ$, 
$T_\oplus \neq T$. Instead, $T_\oplus$ is 
related to $T$ by an offset given by~\cite{Ding:2016lwt}
\begin{equation}
T-T_\oplus = \frac{\lambda_0-\lambda}{360^\circ} T_\text{sidereal} = 3.748\;\text{h} \;,
\label{eq:Toplus}
\end{equation}
where $T_\text{sidereal} = 
23\;\text{h}\;56\;\text{min}\;4.091\;\text{s}$ is the sidereal day.
The first occurrence for which $T_\oplus = 0$
is therefore approximately 
3.75 hours after the 2000 vernal equinox,
or March 20, 2000 at 11:20 UTC. 
This is the chosen 
initial condition for referencing 
the time of ZEUS $e^\pm p$ events.
The co-latitude of HERA is $\chi \approx 36.4^\circ$,
and the electron/positron-beam orientation for ZEUS 
is $\Psi \approx 20^\circ$ south of west.
Finally, the net rotation $R$ 
from the proton-beam direction in the laboratory frame
to the SCF is given by 
\begin{equation}
R = 
\begin{pmatrix}1 & 0 & 0 \\ 0 & 0 & 1 \\ 0 & -1 & 0\end{pmatrix} 
\begin{pmatrix}\cos\Psi & \sin\Psi & 0 \\ -\sin\Psi & \cos\Psi & 0 \\ 
0 & 0 & 1\end{pmatrix} 
\begin{pmatrix}\cos\chi\cos\omega_{\oplus} T_{\oplus} 
& \cos\chi\sin\omega_{\oplus} T_{\oplus} 
& -\sin\chi \\ -\sin\omega_{\oplus} T_{\oplus} 
& \cos\omega_{\oplus} T_{\oplus} & 0 \\ 
\sin\chi\cos\omega_{\oplus} T_{\oplus} 
& \sin\chi\sin\omega_{\oplus} T_{\oplus} & \cos\chi\end{pmatrix},
\label{eq:rotation}
\end{equation}
where 
$\omega_{\oplus} = 2\pi/T_\text{sidereal}$
is the Earth's sidereal frequency.
Performing the rotation $R$ to express 
the laboratory-frame coefficients in terms of the 
constant SCF coefficients induces a sidereal-time dependence
at multiples of the Earth's sidereal frequency.

As an example, the transformation of the coefficient $c_u^{33}$ reads:
\begin{eqnarray}
c_u^{33} & = &\frac{1}{2} (c_u^{XX}+c_u^{YY}) 
    \left(\cos^2\chi \sin^2\Psi +\cos^2\Psi \right) 
    +c_u^{ZZ} \sin^2\chi  \sin^2\Psi  \nonumber \\
&&-2c_u^{XZ}\sin\chi\sin\Psi 
    \left[\cos\chi\sin\Psi\cos(\omega_\oplus T_\oplus)+\cos\Psi\sin(\omega_\oplus T_\oplus)\right]\nonumber\\
&&-2c_u^{YZ}\sin\chi\sin\Psi 
    \left[\cos\chi\sin\psi\sin(\omega_\oplus T_\oplus)-\cos\psi\cos(\omega_\oplus T_\oplus)\right]\nonumber\\
&&+c_u^{XY}
    \left[(\cos^2\chi\sin^2\Psi-\cos^2\Psi)\sin(2\omega_\oplus T_\oplus) - \cos\chi\sin(2\Psi)\cos(2\omega_\oplus T_\oplus)\right] \\
&&+\frac{1}{2} (c_u^{XX}-c_u^{YY}) 
    \left[(\cos^2\chi\sin^2\Psi-\cos^2\Psi)\cos (2\omega_\oplus T_\oplus) + \cos\chi\sin(2\Psi)\sin(2\omega_\oplus T_\oplus)\right]\nonumber,
\end{eqnarray}
where it is seen that the terms 
proportional to 
$(c_u^{XX}+c_u^{YY})$ and $c_u^{ZZ}$ possess no 
sidereal-time dependence, 
the terms proportional to $c_u^{XZ}$ and $c_u^{YZ}$ 
oscillate with angular frequency $\omega_\oplus$ and the 
terms proportional to $c_u^{XY}$ and $(c_u^{XX}-c_u^{YY})$ 
oscillate with angular frequency $2 \omega_\oplus$.

After expressing the cross 
section given in Eq.~\eqref{xsecc}
in terms of the SCF coefficients, 
only the following 18 combinations 
of coefficients yield sidereal-time oscillations:
\begin{equation}
c_{f}^{TX} \;, c_{f}^{XZ} \;,
c_{f}^{TY} \;, c_{f}^{YZ}  \;,
c_{f}^{XY} \; \text{and} \; (c_{f}^{XX}-c_{f}^{YY}) \; ,
\label{cSCF}
\end{equation}
with $f=u$, $d$, and $s$. Analogously, 
after the replacement $a_\text{lab}^{(5)\mu\alpha\beta} 
= \Lambda^\mu_{\phantom{\mu}\nu}\Lambda^\alpha_{\phantom{\alpha}\rho}
\Lambda^\beta_{\phantom{\beta}\sigma}a_\text{SCF}^{(5)\nu\rho\sigma}$ 
in Eq.~\eqref{xseca5}, it is found that, for each quark, only the 
following 12 combinations of coefficients 
yield sidereal-time oscillations:
\begin{eqnarray}
&&
(a_{\text{S}f}^{(5)TXX}-a_{\text{S}f}^{(5)TYY}) ,\; 
(a_{\text{S}f}^{(5)XXZ}-a_{\text{S}f}^{(5)YYZ}) ,\; 
a_{\text{S}f}^{(5)TXY} ,\; 
a_{\text{S}f}^{(5)TXZ} ,\; 
a_{\text{S}f}^{(5)TYZ} ,\; 
a_{\text{S}f}^{(5)XXX} ,\; \nonumber \\
&&
a_{\text{S}f}^{(5)XXY} ,\; 
a_{\text{S}f}^{(5)XYY} ,\; 
a_{\text{S}f}^{(5)XYZ} ,\; 
a_{\text{S}f}^{(5)XZZ} ,\; 
a_{\text{S}f}^{(5)YYY} \; \text{and} \; 
a_{\text{S}f}^{(5)YZZ} ,
\label{a5SCF}
\end{eqnarray}
giving a total of 36 combinations. For antiquarks with 
flavour $f = \bar u, \bar d$ and $\bar s$, 
the coefficients effectively 
appear in the cross section Eq.~\eqref{xseca5}
with opposite signs relative 
to the quark coefficients. Since the contributions from $s$ and $\bar s$ are equal, 
only the 24 coefficients for $f = u$ and $d$
are considered. 
The $c$- and $a^{(5)}$-type coefficients 
in Eqs.~\eqref{cSCF} and \eqref{a5SCF} induce 
oscillations with frequencies up to 
$2\omega_\oplus$ and $3\omega_\oplus$, 
respectively. More generally, an SCF 
coefficient with $n$ Lorentz indices 
will include sidereal oscillations up to $n\omega_\oplus$.

\begin{figure}[p]
\begin{center}
\includegraphics[width=0.965 \linewidth]{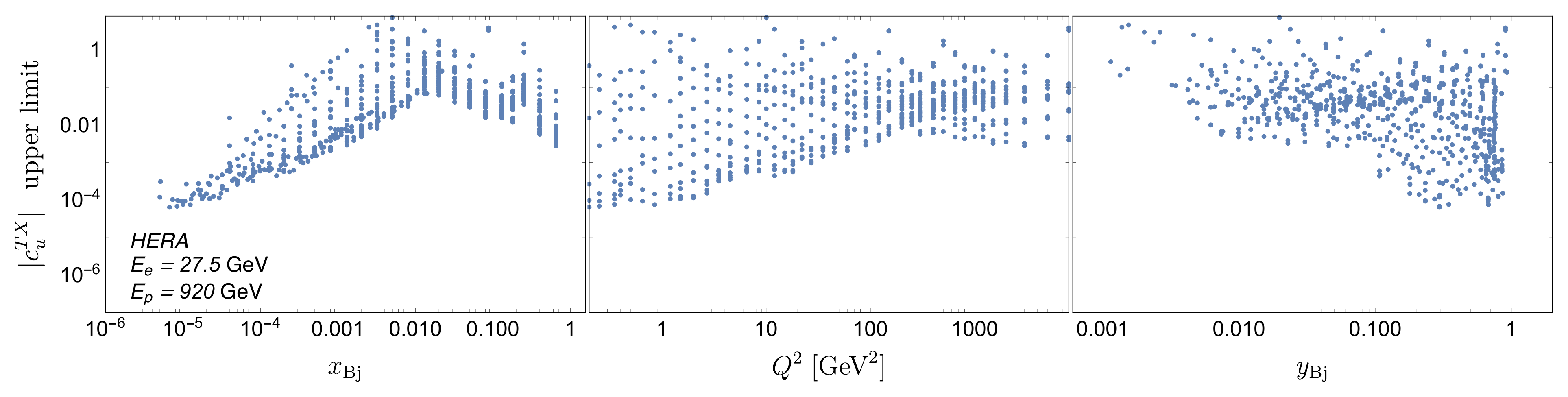}
\includegraphics[width=0.99 \linewidth]{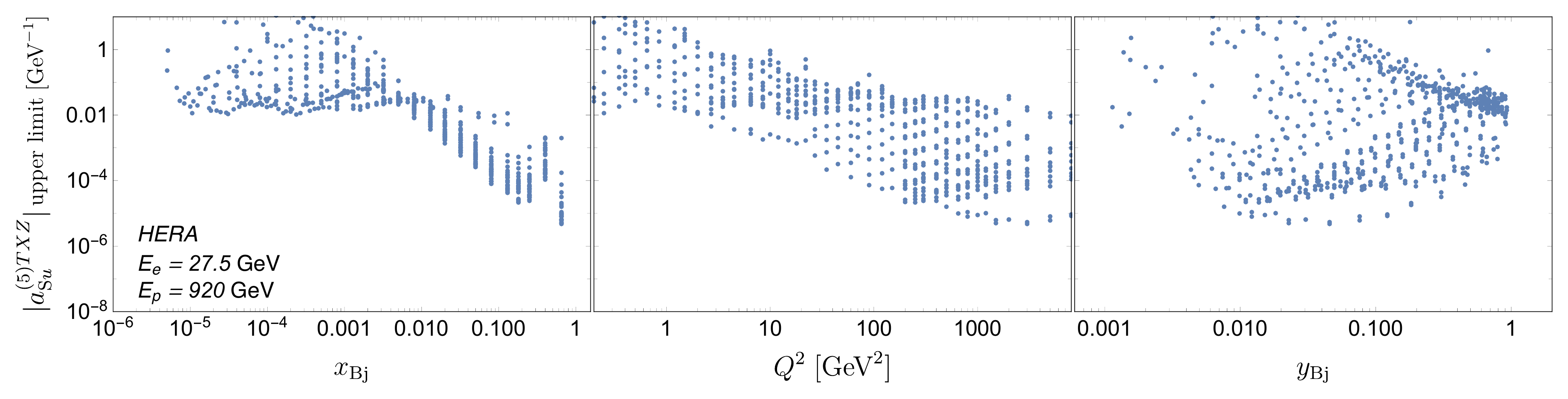}
\end{center}
\caption{Regions in $x_{\textnormal{Bj}}$, 
$Q^2$ and $y_{\textnormal{Bj}}$ that have 
sensitivity to Lorentz-violating effects for a 
single $c$-type and $a^{(5)}$-type coefficient. 
The points displayed in the plots are 
taken from Figures 2 and 4 of Ref.~\protect\cite{klsv19},
where no experimental restrictions on the
kinematic region have been considered.}
\label{fig:c_a5_distribution}
\end{figure}

The expected sensitivities to the $c$- 
and $a^{(5)}$-type coefficients that 
can be extracted from DIS data 
have been studied~\cite{klv17, ls18, klsv19}. 
In particular, the combined DIS 
cross sections from ZEUS and H1~\cite{hera}
have been used to estimate the 
sensitivity of a sidereal-time
study using these data.
For each value in 
the $(x_{\text{Bj}},Q^2)$ plane, 
potential constraints based on four 
sidereal intervals have been extracted. 
A subset of simulated results describing 
the sensitivity to the $c$- and 
$a^{(5)}$-type coefficients as 
a function of phase are shown in 
Fig.~\ref{fig:c_a5_distribution}.
It is observed that the sensitivities to 
the $c$- and $a^{(5)}$-type coefficients 
are roughly at levels of $10^{-4}$ 
and $10^{-6}$\;GeV$^{-1}$, respectively.

\section{Experimental set-up and data selection}
\label{sec-exp}
The analysis is based on events collected 
with the ZEUS detector at HERA during the 
HERA II run period 2003--2007.
For this configuration, the initial-state
proton- and electron/positron-beam energies 
were $E_p=920$\;GeV and $E_e = 27.5$\;GeV, 
respectively, with 
a centre-of-mass energy of 
$\sqrt{s} = 318$\;GeV and
an integrated luminosity 
372\;pb$^{-1}$. 
Details of the ZEUS detector 
are given elsewhere~\cite{Holm:1993frx}.
%

The NC DIS events were selected with the following criteria
~\cite{ZEUS:2019jya}:
\begin{itemize}
\item 
the final-state lepton was identified using an
algorithm based on a 
neural network~\cite{Abramowicz:1995zi, Sinkus:1996ch},
giving a probability larger than 90\%;
\item
the energy of the final-state lepton 
$E'_e > 10$ GeV to ensure a 
high electron-identification efficiency;
\item $Q^2 > 5\; \text{GeV}^2$;
\item $\theta_e > 1\; \text{rad}$, where
$\theta_e$ is the scattering angle 
between the outgoing lepton and incoming 
proton direction to ensure the high efficiency
of the electron-identification algorithm. 
This provides an upper limit on $Q^2$;
\item
the scattered lepton was required to enter the 
calorimeter at a radial position 
larger than $15\; \text{cm}$, implying an 
upper bound on the lepton 
scattering angle $\theta_e \lesssim 3 \; \text{rad}$;
\item the position of the event vertex along
the laboratory $z$ axis was required to
be within $30$ cm of its 
nominal value and the transverse distance
of the event vertex from the interaction
point was required to be within $0.5$ cm, to reject background;
\item 
$47 \; \GeV < E - p_z < 69 \; \GeV$,
where $E$ and $p_z$ are the total energy and $z$-component
of the final state, to reject background.
\end{itemize}
This selection resulted in $4.5\cdot 10^7$ 
events covering the kinematic range 
$7.7\cdot 10^{-5} < x_{\text{Bj}} <1$ and $5 < Q^2 < 8800 \; \GeV^2$.
%
\section{Analysis method}
\label{ssec:observables}
Possible variations of cross sections with periodicity
$T_{\text{p}}$ were studied. 
The starting point was the triple differential DIS cross section
\begin{equation}
\frac{d\sigma}{dx_{\text{Bj}}\; dQ^2 \; d\phi_{T_\text{p}}}\;,
\label{eq:startanalysis}
\end{equation}
where the temporal
phase $\phi_{T_{\text{p}}} 
= \text{Mod}(T_\oplus,T_{\text{p}})/T_{\text{p}}$ is 
the phase of a given DIS event with
the time stamp $T_\oplus$ 
for the chosen period, $T_{\text{p}}$,
and is defined in the range $[0,1]$. Within the SME, 
only $T_{\text{p}} = T_{\text{sidereal}}$
yields a non-vanishing dependence 
on $\phi_{T_{\text{p}}}$,
as the sidereal angle is $\omega_\oplus T_\oplus$.
If a different period is used, the 
time dependence 
quickly averages out; 
the larger the difference between 
$T_{\text{p}}$ and $T_{\text{sidereal}}$, 
the faster this occurs.

It was necessary to eliminate uncertainties related to the 
instantaneous luminosity, which decays over
several hours during each fill. To do so would have required
the integrated luminosity to be measured roughly every minute,
but this information was not available.
Instead, double ratios of the form 
%
\begin{equation}
r({\text{PS}}_1, {\text{PS}}_2) = 
\frac{
\int_{{\text{PS}}_1 } dx_{\text{Bj}} 
dQ^2 \frac{d\sigma}{dx_{\text{Bj}}\; dQ^2 \; d\phi_{T_{\text{p}}}}
/
\int_{{\text{PS}}_1 } dx_{\text{Bj}} dQ^2 d\phi_{T_{\text{p}}} \frac{d\sigma}{dx_{\text{Bj}}\; dQ^2 \; d\phi_{T_{\text{p}}}
}}{
\int_{{\text{PS}}_2 } dx_{\text{Bj}} dQ^2 \frac{d\sigma}{dx_{\text{Bj}}\; dQ^2 \; d\phi_{T_\text{p}}}
/
\int_{{\text{PS}}_2 } dx_{\text{Bj}} dQ^2 d\phi_{T_{\text{p}}} \frac{d\sigma}{dx_{\text{Bj}}\; dQ^2 \; d\phi_{T_{\text{p}}}
}}\;,
\label{r}
\end{equation}
where ${\text{PS}}_1$ and ${\text{PS}}_2$ are 
two regions of the $(x_{\text{Bj}},Q^2)$ kinematic range, for which luminosity uncertainty cancels, were used. 
Essential properties of Eq.~\eqref{r} are 
that it is independent of the luminosity 
and equal to unity in the absence of SME effects.
Evaluating the 
contributions involving integrations 
of $\phi_{T_{\text{p}}}$ 
results in zero sensitivity
to SME coefficients and 
when summing the contributions 
of all quark flavours produces
the Standard Model cross section.
Moreover, if the regions ${\text{PS}}_{1,2}$ 
are chosen appropriately, the double ratio shows a 
strong $\phi_{T_{\text{p}}}$ dependence in the presence 
of SME effects.
As explained in Section~\eqref{ssec:DIS}, a strong sensitivity at 
low $x_{\text{Bj}}$ ($\approx 10^{-5}$--$10^{-3}$) for the 
$c$-type coefficients and 
$x_{\text{Bj}} \gtrsim 10^{-3}$
for the $a^{(5)}$-type coefficients is expected. 
However, for $x_{\text{Bj}} \lesssim 10^{-4}$ 
and $x_{\text{Bj}} \gtrsim 10^{-2}$,
fewer DIS events are available
and thus the statistical uncertainties become larger. 
The $x_{\text{Bj}}$ distribution is sharply peaked
around $x_{\text{Bj}}\approx 10^{-3}$, implying a statistically 
optimal choice of cut $x_{\text{c}}$ 
in the $(x_{\text{Bj}},Q^2)$ kinematic range.

In evaluating Eq.~\eqref{r} for 
several values of $x_{\text{c}}$, it was found 
that the combinations simultaneously giving the 
most sensitivity to the cross section while 
producing the lowest statistical uncertainties
are 
$x_{\text{c}}=5\cdot 10^{-4}$ and 
$10^{-3}$ for $c$-
and $a^{(5)}$-type coefficients, respectively.
However, as no higher-order QCD corrections are included 
in the cross sections given in Eqs.~\eqref{xsecc} and \eqref{xseca5},
potential large corrections are expected, 
of the order $[\alpha_s (Q^2) \log 1/x_{\text{Bj}}]^n$ at very 
low $x_{\text{Bj}}$, which need to be resummed.  Numerically,
such effects could shift the value of $x$ entering
the PDFs. Thus, $x$ would not coincide 
with $x_{\text{Bj}}$.
This issue is partially alleviated by cutting
at moderate values $x_{\text{Bj}}\gtrsim 10^{-3}$, where
QCD corrections have a smaller influence. The value
$x_{\text{c}} = 10^{-3}$ was therefore chosen for both 
the $c$- and $a^{(5)}$-type coefficients. 

By employing Eqs.~\eqref{xsecc} and \eqref{xseca5},
the integrations above and below 
$x_{\text{c}} = 10^{-3}$
were performed for the kinematic
range in $x_{\text{Bj}},Q^2$ and $\theta_e$ as 
described in Section~\ref{sec-exp}.
The double ratios from Eq.~\eqref{r} evaluated in the 
laboratory frame are:
\begin{eqnarray}
r_{c}(x_{\text{Bj}}> x_{\text{c}} , x_{\text{Bj}}< x_{\text{c}} ) & = & 
1  - 12.8 \; c_u^{03}  - 13.9 \; c_u^{33} + 0.9 \;(c_u^{11} +  c_u^{22})\nonumber \\
&&\phantom{1}- 4.2 \; c_d^{03}  -  2.9 \; c_d^{33} + 0.1 \; (c_d^{11} +  c_d^{22})\nonumber \\
&&\phantom{1}- 3.4 \; c_s^{03}  -  1.8 \; c_s^{33} + 2.9\cdot 10^{-2} \; (c_s^{11} +  c_s^{22}) \; ,
\label{eq:rc}\\
r_{a^{(5)}}(x_{\text{Bj}}>x_{\text{c}} , x_{\text{Bj}} < x_{\text{c}} ) & = & 
1 
- 6.1 \cdot 10^3 \; a_u^{(5)003} + 6.8 \cdot 10^3 \;  a_u^{(5)033}  - 2.5\cdot 10^3 \;  a_u^{(5)333} \nonumber   \\
&& \phantom{1 } + 5.0\cdot 10^2 \;  (a_u^{(5)113} + a_u^{(5)223} - a_u^{(5)011}  - a_u^{(5)022}) \nonumber   \\
&&
\phantom{1} - 4.1 \cdot 10^2 \;  a_d^{(5)003} +  4.7\cdot 10^2  \; a_d^{(5)033}
 - 1.7 \cdot 10^2  \; a_d^{(5)333}\nonumber \\
&& \phantom{1 } + 40  \; (a_d^{(5)113 } + a_d^{(5)223} - a_d^{(5)011} - a_d^{(5)022} ) \; ,
\label{eq:ra5}
\end{eqnarray}
where the numerical prefactors 
multipling the $a^{(5)}$-type coefficients
in the latter
expression are in units of GeV.
These expressions only contain terms 
linear in the coefficients,
as higher-order corrections are greatly suppressed.
The ratio $r_{a^{(5)}}$ has negligible
sensitivity to $a_s^{(5)\mu\alpha\beta}$
because of the nearly identical and opposite-in-sign 
$s, \bar s$ contributions
to the cross section, see Eq.~\eqref{xseca5}.
After re-expressing 
the laboratory-frame coefficients appearing
in Eqs.~\eqref{eq:rc} and \eqref{eq:ra5} 
in terms of 
rotational-symmetry violating combinations of 
the SCF coefficients described
in Section~\ref{ssec:sidsignals} 
using Eq.~\eqref{eq:rotation},
it is found that 42 SCF coefficients
result in deviations of 
$r_{c}$ or $r_{a^{(5)}}$ from unity. 
Comparisons between these double ratios 
and the analogous ratios constructed
purely from the binned DIS events 
were made under the assumption of constant
efficiency corrections.
Constraints were placed on each coefficient
one at a time by setting all 
others to zero, in accordance
with standard practice~\cite{tables}. 
\section{Systematic effects}
\label{ssec:syst}
\subsection{Initial considerations}
\label{ssec:initial}
%
A luminosity-independent ratio, Eq.~\eqref{r},
that does not possess 
a sizeable dependence on SME effects was used
to test the data. 
The ratio $r(Q^2>Q_{\text{c}}^2,Q^2<Q^2_{\text{c}})$ 
with $Q_{\text{c}}^2 = 20$\;GeV$^2$
fulfills this requirement. 

\begin{figure}[p]
\begin{center}
 \includegraphics[width=0.49 \linewidth]{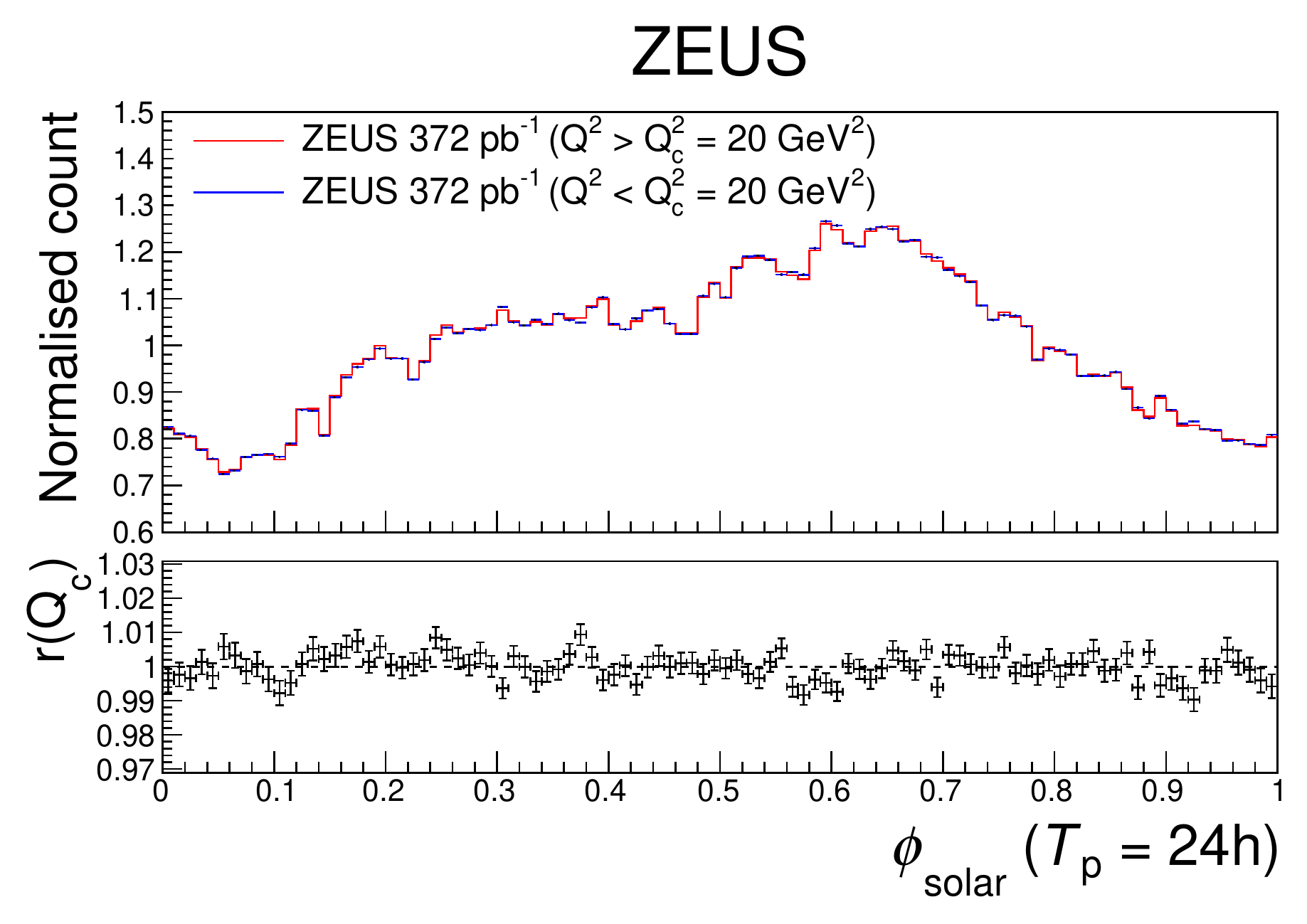}
 \includegraphics[width=0.49 \linewidth]{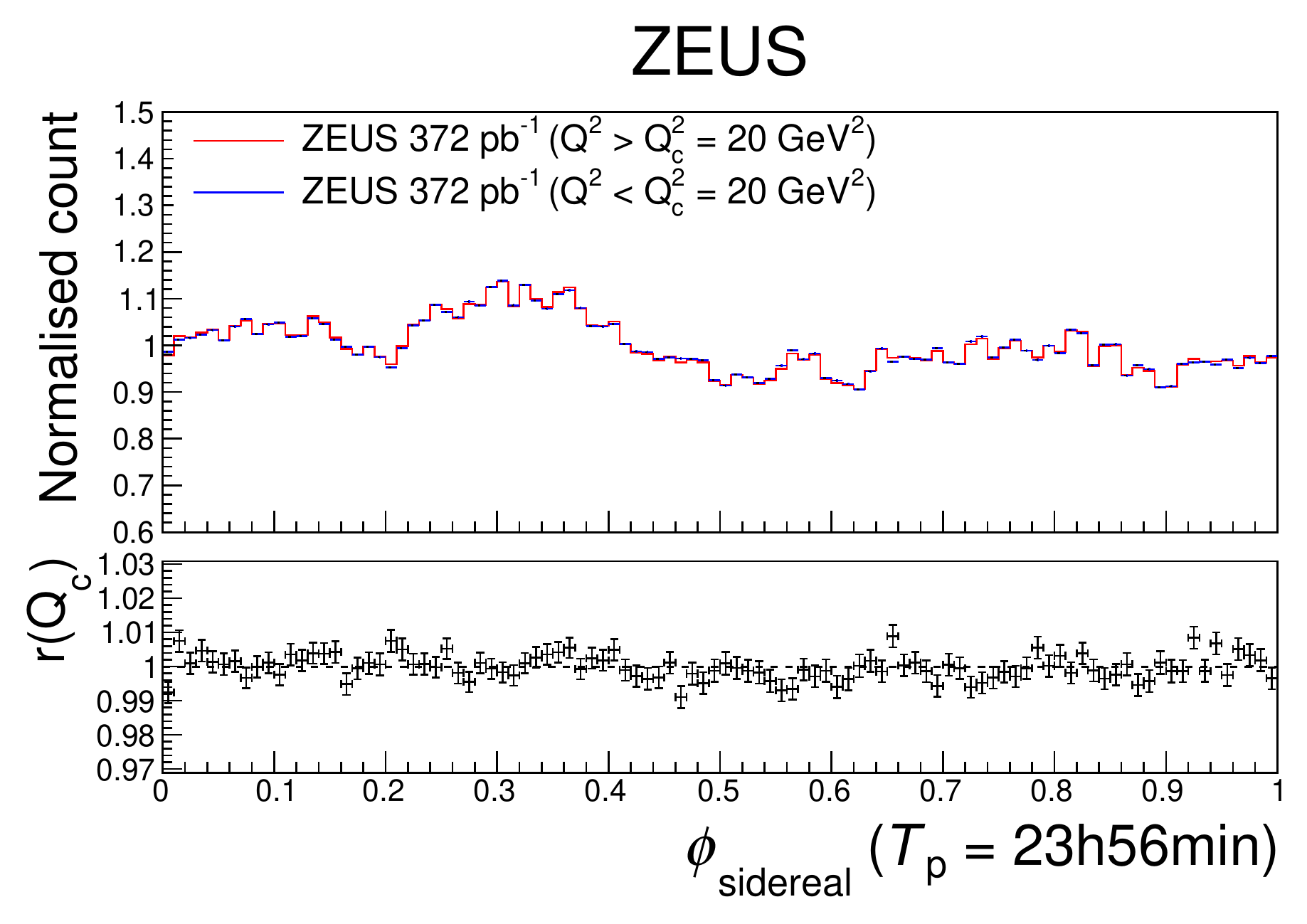} \\
 {\scriptsize (a)} \hskip 8cm {\scriptsize (b)} \\ \; \\
 \includegraphics[width=0.49 \linewidth]{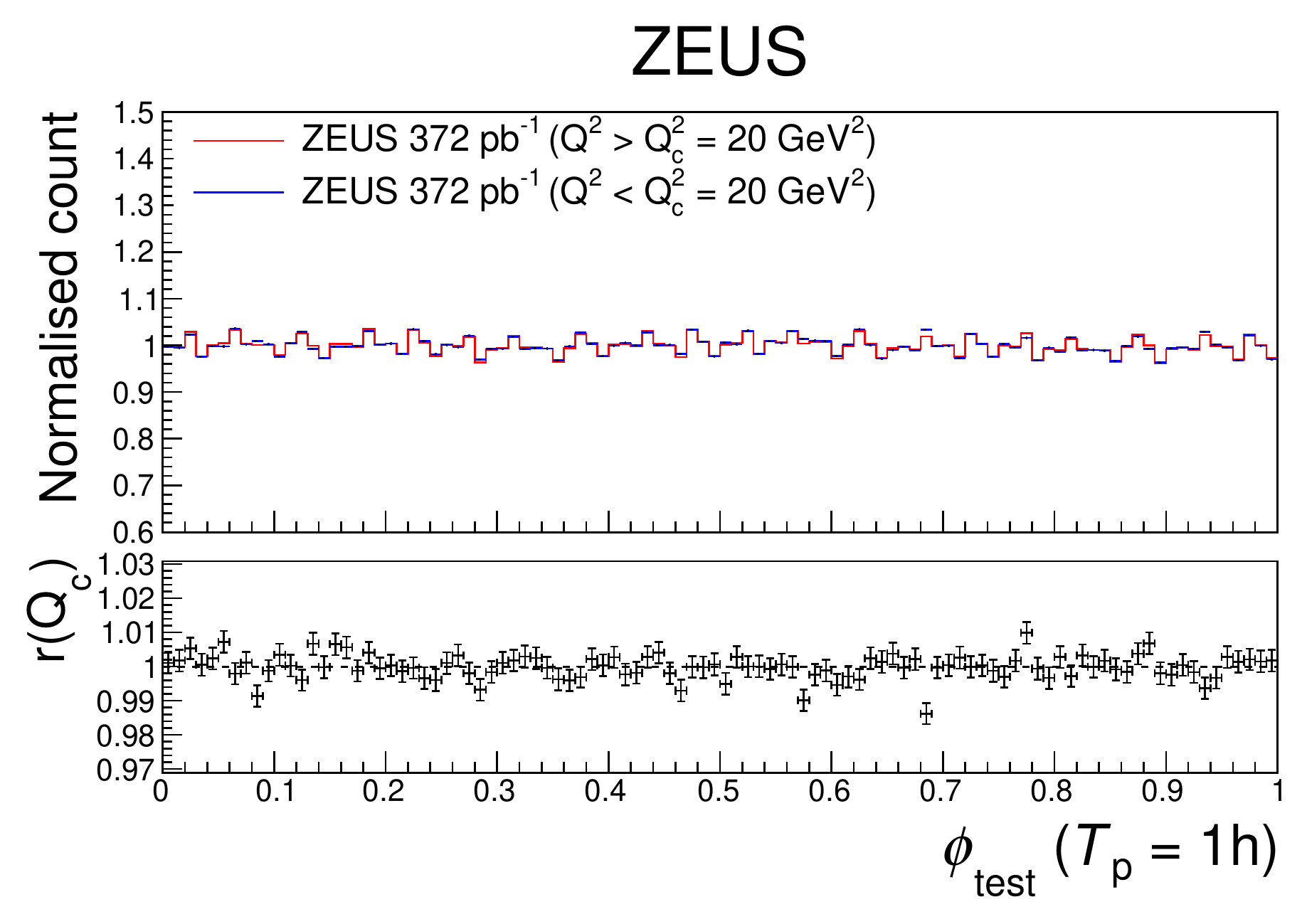}
 \includegraphics[width=0.49 \linewidth]{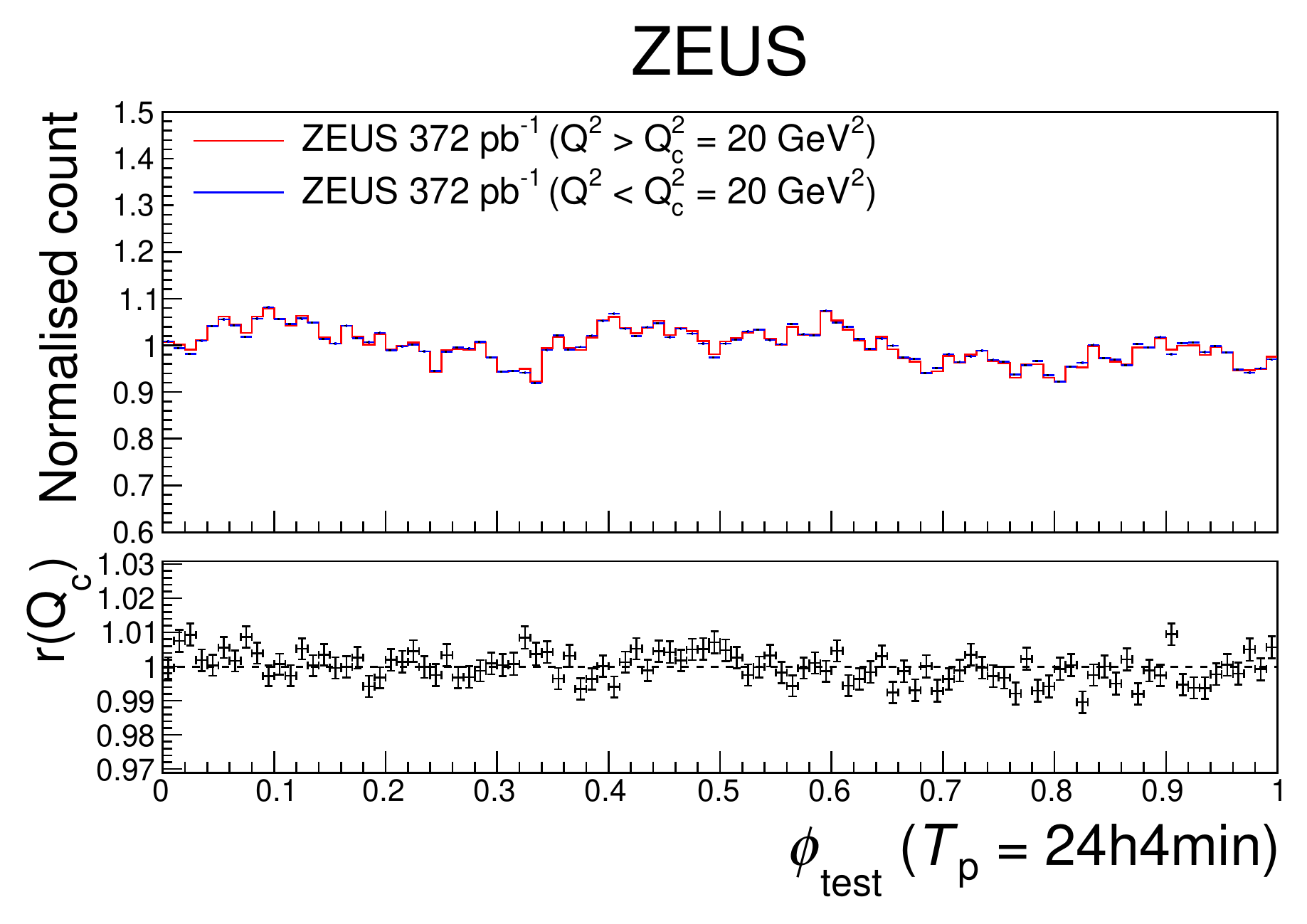}\\
 {\scriptsize (c)} \hskip 8cm {\scriptsize (d)}
\end{center}
\caption{
Solar (a), sidereal (b), $T_{\textnormal{p}}=1$\;\textnormal{h} (c) and 
$T_{\textnormal{p}}=24$\;\textnormal{h}\;$4$\;\textnormal{min} (d)
phase dependence of the normalised counts in 100 bins
with the kinematic region divided
by $Q_{\textnormal{c}}^2 = 20$\;\textnormal{GeV}$^2$. 
The vertical axis displays the number of events
per bin normalised to the total number
of events times the bin width. The ratios 
of the counts $r(Q_{\textnormal{c}})$ above and
below $Q_{\textnormal{c}}^2$ are given in the bottom panels.
For the solar phase, $\phi_{\textnormal{solar}} = 0$ 
is identified with 11:20 UTC. 
Only statistical uncertainties are displayed.
}
\label{fig:DIS-20}
\end{figure}

The data rate in the detector is not constant.
In particular, more data were taken in the evenings
than in the morning.
To display this effect, 
the entire DIS selection 
in the two $Q^2$ regions 
was binned using the solar phase 
$\phi_\text{solar} = 
\text{Mod}(T_\oplus,T_\text{solar})/T_\text{solar}$ 
with $T_\text{solar} =24$\;h.
In Fig.~\ref{fig:DIS-20}(a), 
the resulting normalised count 
of the events passing the selection 
described in Section~\ref{sec-exp} 
is displayed.
It is clear that 
an $\mathcal{O}(25\%)$ effect is present. 
Switching from solar to sidereal binning 
with $\phi_{\text{sidereal}} 
= \text{Mod}(T_\oplus, T_{\text{sidereal}})/T_\text{sidereal}$
dilutes this solar-phase dependence 
for long data-taking periods. 
As shown in Fig.~\ref{fig:DIS-20}(b),
sidereal dilution over $\approx 5$ years 
reduces the effect to $\mathcal{O}(10\%)$,
but is not sufficient to erase it completely.
The dilution effect of a much longer period of 
data taking can be simulated by binning using 
a short period. In Fig.~\ref{fig:DIS-20}(c), a plot of the same 
data binned using a $T_{\text{p}}=1$\;h period 
with $\phi_{\text{test}} 
= \text{Mod}(T_\oplus, 1\;\text{h})/1\;\text{h}$
is shown. The initial solar-phase 
dependence disappears almost completely.
Figure~\ref{fig:DIS-20}(d) shows that
for a period slightly longer than the
solar day, $T_{\text{p}}=24$\;h\;4\;min,
a similar level of dilution as for 
the sidereal-phase
binning in Fig.~\ref{fig:DIS-20}(b) is observed.
For each choice of phase, the
histograms for events 
with $Q^2$ above and below $Q_{\text{c}}^2$ 
closely track each other and 
their ratios are consistent with unity. 

\begin{figure}[p]
\begin{center}
\includegraphics[width=0.99 \linewidth]{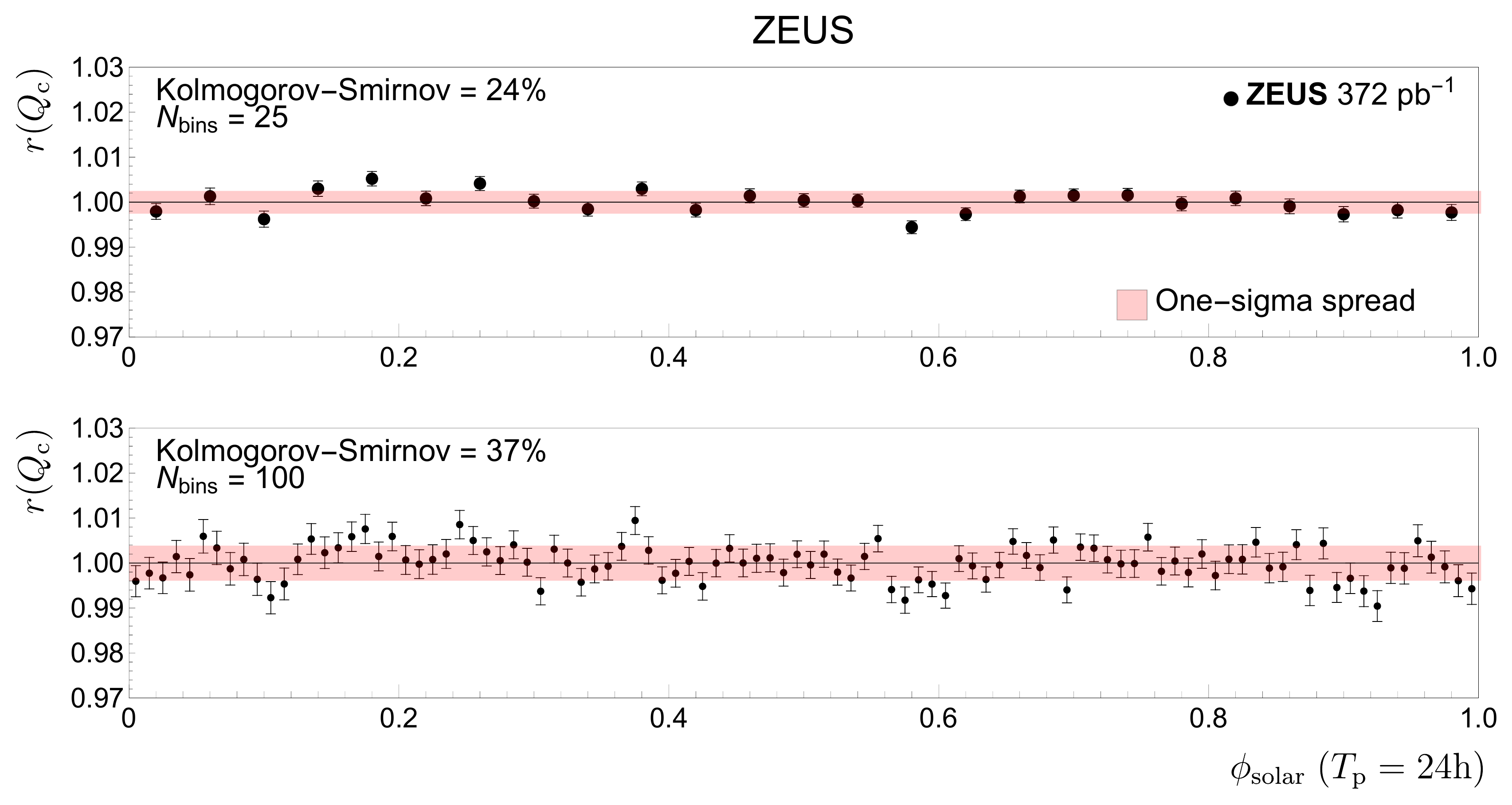}
\end{center}
\caption{
Ratio $r(Q_{\textnormal{c}})$ of normalised 
counts for $Q_{\textnormal{c}}^2 = 20$\;\textnormal{GeV}$^2$ 
binned with the solar 
phase where $N_{\textnormal{bins}} = 25$ and $100$. 
The case $N_{\textnormal{bins}} = 100$ from
Fig.~\ref{fig:DIS-20}(a) is repeated.
The displayed uncertainties include 
statistical uncertainties only and
the one-sigma spreads (bands)
are the standard deviations of the 
central values.
The observed distributions are
compared to a Gaussian distribution in which 
only statistical errors are included 
using a Kolmogorov--Smirnov test.
}
\label{fig:KS-solar-20-5-8}
\end{figure}

\begin{figure}[p]
\begin{center}
\includegraphics[width=0.99 \linewidth]{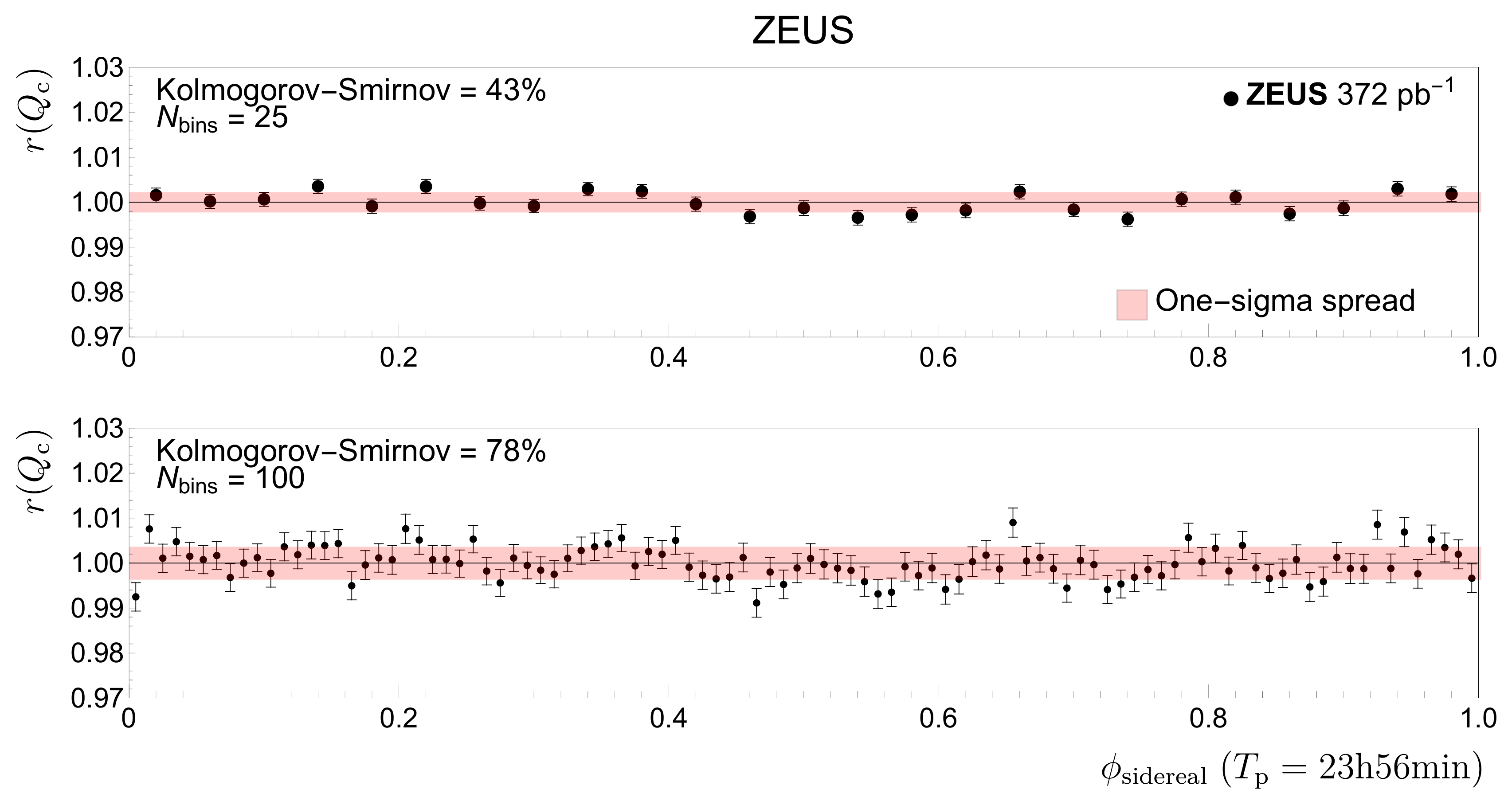}
\end{center}
\caption{
Ratio $r(Q_{\textnormal{c}})$ of normalised 
counts for $Q_{\textnormal{c}}^2 = 20$\;\textnormal{GeV}$^2$ 
binned with the sidereal
phase where $N_{\textnormal{bins}} = 25$ and $100$. 
The displayed uncertainties include 
statistical uncertainties only and
the one-sigma spreads (bands)
are the standard deviations of the 
central values.
The observed distributions are
compared to a Gaussian distribution in which 
only statistical errors are included 
using a Kolmogorov--Smirnov test.
}
\label{fig:KS-sidereal-20-5-8}
\end{figure}

An important question is whether systematic 
uncertainties can be neglected.
A partial answer can be found by performing 
Kolmogorov--Smirnov (KS) tests on 
the binned ratios $r(Q_{\text{c}})$.
This test calculates the probability that the 
observed distributions are compatible with an unsorted 
sampling of a normal distribution with mean unity 
and standard deviation identical to the 
observed statistical uncertainties. 
The ratios are plotted in different numbers of
bins $N_{\text{bins}}$ to estimate the size of systematic 
relative to statistical effects. 
The results are presented for 
$T_{\text{p}} = 24$\;h in Fig.~\ref{fig:KS-solar-20-5-8}
and for $T_{\text{p}} = 
T_{\text{sidereal}}$ in Fig.~\ref{fig:KS-sidereal-20-5-8}.
It is observed that, 
independently of the 
period, $T_{\text{p}} = 24$\;h 
or $T_{\text{p}} = T_{\text{sidereal}}$, and 
of the number of bins,
the results are statistically 
compatible with unity 
as indicated by the high KS probabilities.
The impact of systematic 
uncertainties appears to be minimal
in these distributions.
However, this is insufficient to
conclude that systematic uncertainties can 
be neglected when looking at sidereal and solar 
distributions for SME-sensitive $x_{\text{c}}$ 
ratios given by Eqs.~\eqref{eq:rc} and \eqref{eq:ra5}. 
This is because low- and high-$x_{\text{Bj}}$ 
regions, where the trigger behaves
differently and could be affected 
by the instantaneous luminosity,
could be subject to systematic effects
which would not cancel in the
ratio defined in Eq.~\eqref{r}.

As long as each bin is considerably smaller 
than the duration of a fill,
which is typically several hours,
(e.g., for $T_{\text{p}}=1$\;h, each 
bin is between $0.6$ min ($N_{\text{bins}}=100$) 
and 2.4 min ($N_{\text{bins}}=25$)),
it is expected that all such effects will average out. 
On the other hand, for the solar or sidereal phase, 
the time bins range between $\approx$ 15 min and 1 h.
It is possible that fluctuations in 
trigger efficiencies and accelerator effects 
could affect high-luminosity and low-luminosity
parts of the fill differently. This might
result in non-negligible residual
effects in the solar- and sidereal-ratio distributions.

\subsection{Data analysis of $\boldsymbol{x_{\text{c}}}$ and Monte Carlo simulation}
\label{ssec:xc_analysis}
\begin{figure}[h]
\begin{center}
\includegraphics[width=0.49 \linewidth]{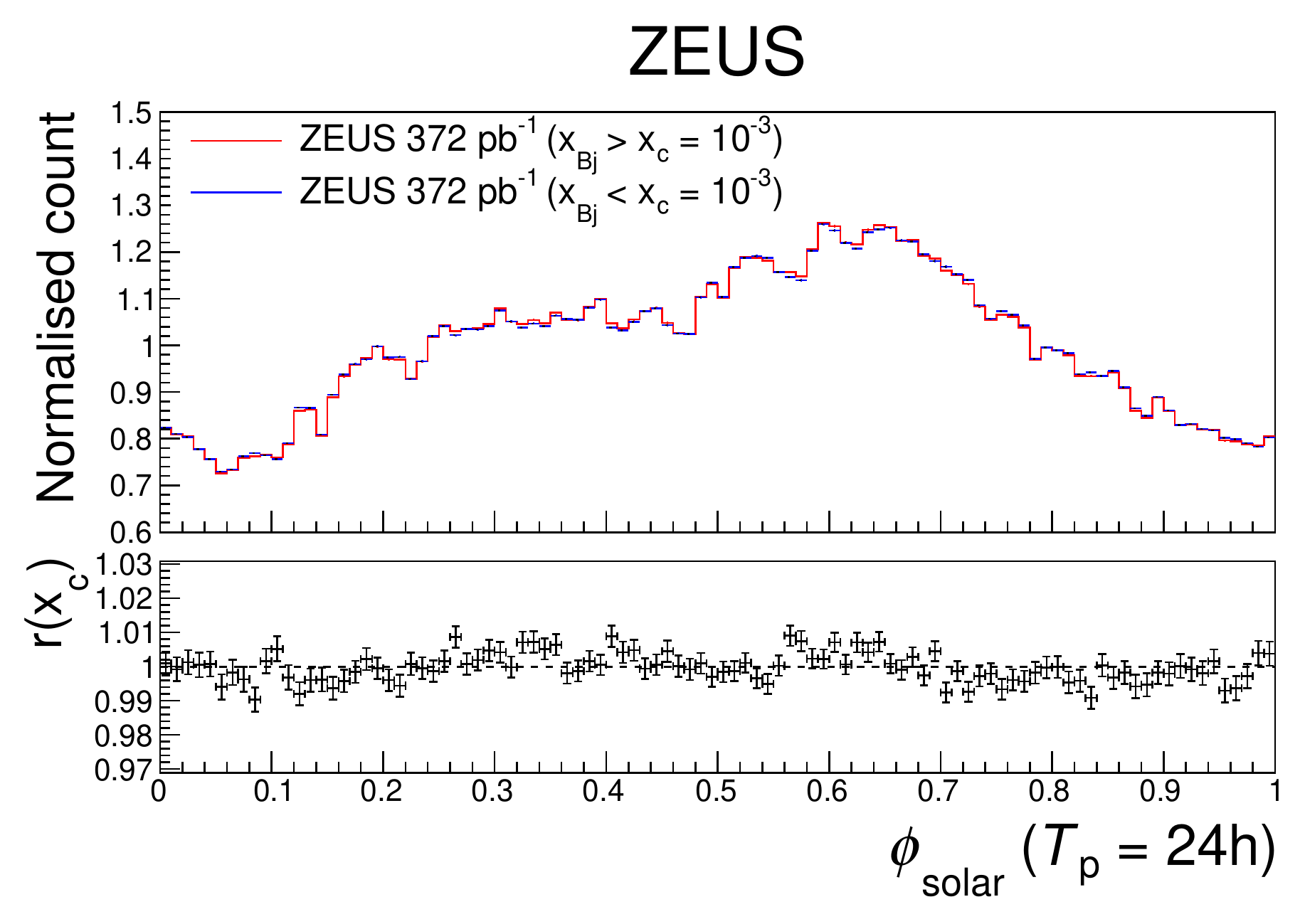}
\includegraphics[width=0.49 \linewidth]{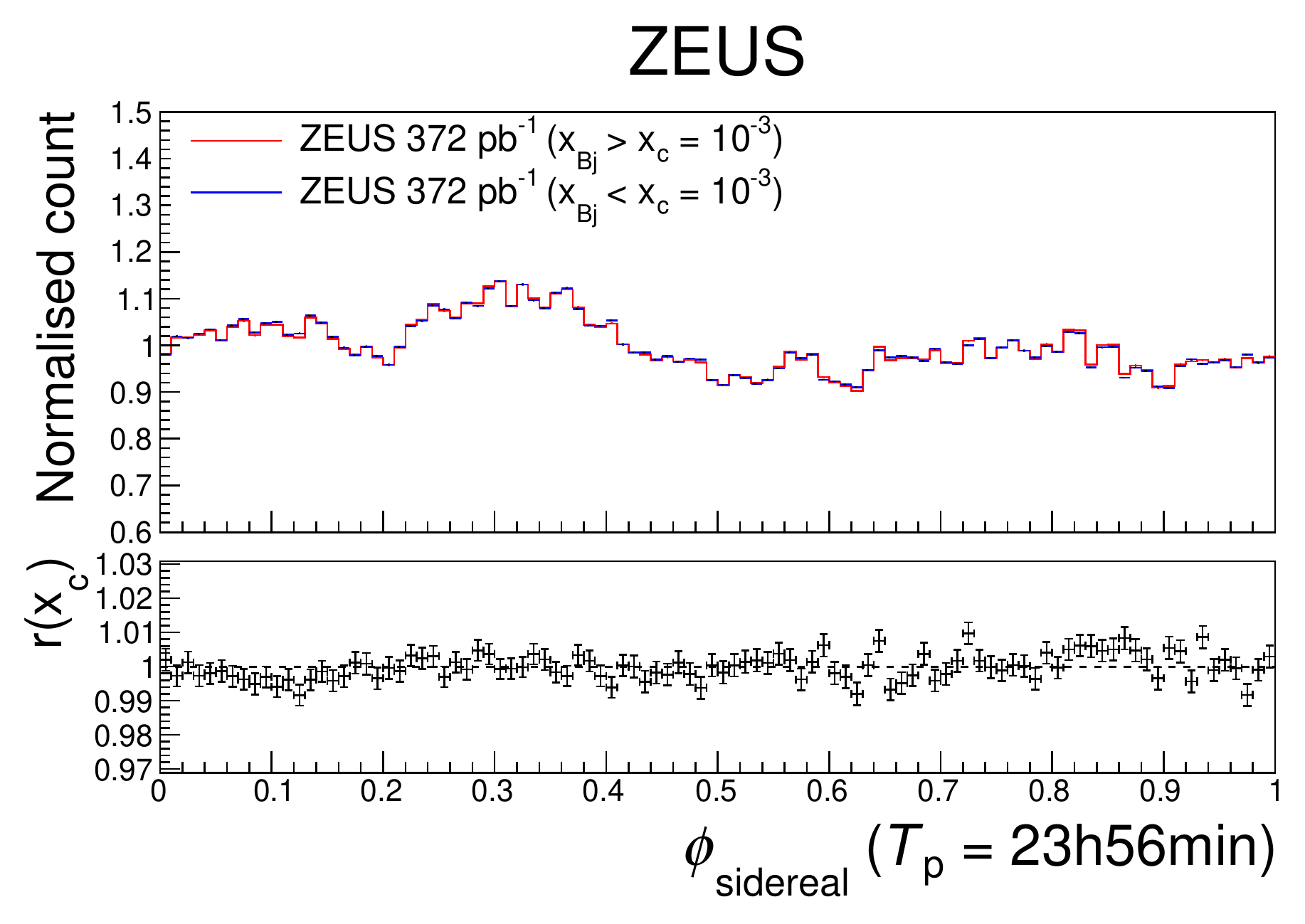}\\
{\scriptsize (a)} \hskip 8cm {\scriptsize (b)} \\ \; \\
\includegraphics[width=0.49 \linewidth]{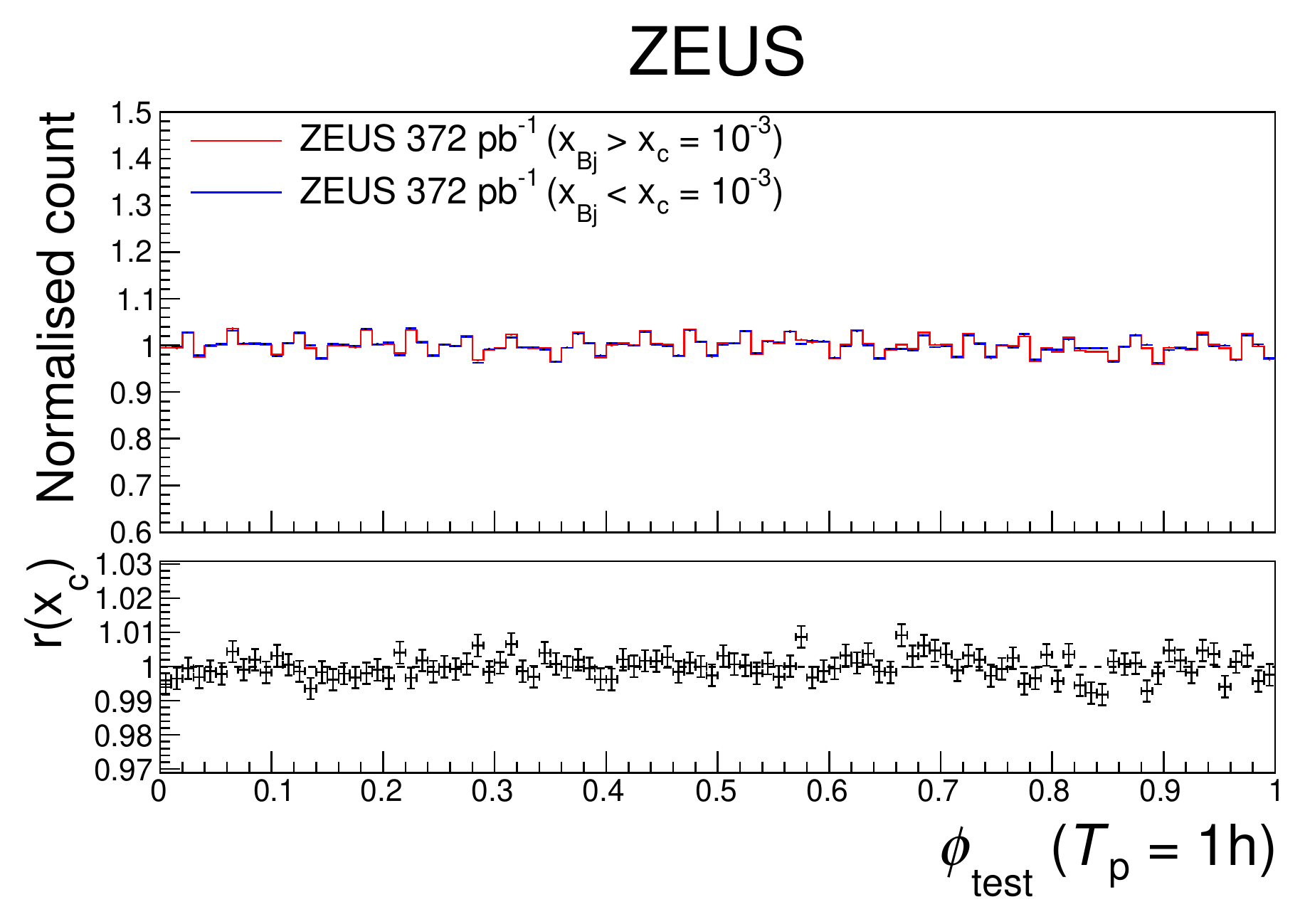}
\includegraphics[width=0.49 \linewidth]{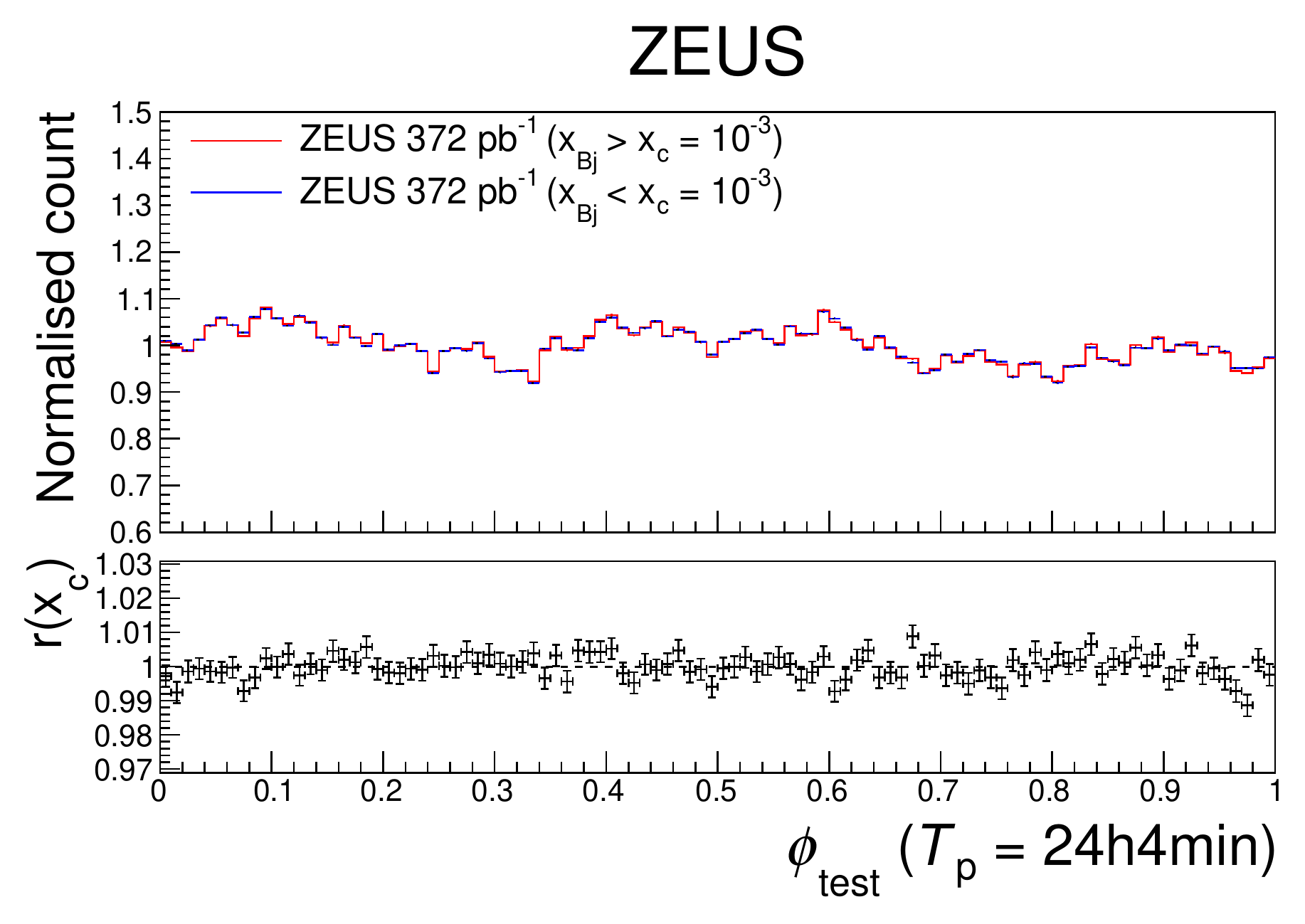}\\
{\scriptsize (c)} \hskip 8cm {\scriptsize (d)}
\end{center}
\caption{
Solar (a), sidereal (b), $T_{\textnormal{p}}=1$\;\textnormal{h} (c) and
$T_{\textnormal{p}}=24$\;\textnormal{h}\;$4$\;\textnormal{min} (d)
phase dependence of the normalised counts in 100 bins
with the kinematic region divided
by $x_{\textnormal{c}} = 10^{-3}$.
The vertical axis displays the number of events
per bin normalised to the total number
of events times the bin width. The ratios
of the counts $r(x_{\textnormal{c}})$ above and
below $x_{\textnormal{c}}$ are given in the bottom panels.
For the solar phase, $\phi_{\textnormal{solar}} = 0$
is identified with 11:20 UTC.
Only statistical uncertainties are displayed.
}
\label{fig:DIS-0.001}
\end{figure}

\begin{figure}[p]
\begin{center}
\includegraphics[width=0.99 \linewidth]{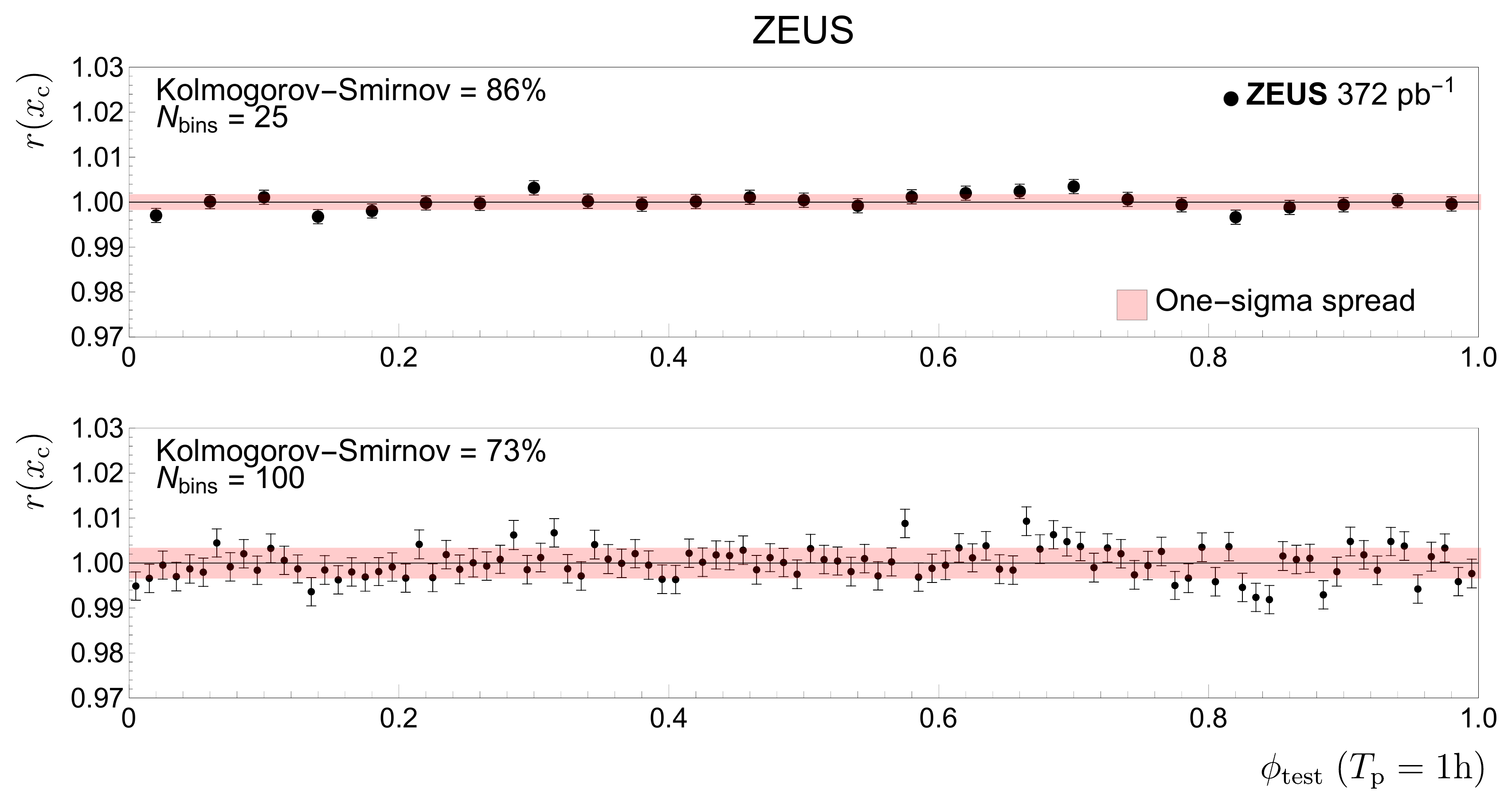}
\end{center}
\caption{
Ratio $r(x_{\textnormal{c}})$ of normalised
counts for $x_{\textnormal{c}} = 10^{-3}$
binned using the test phase with
$T_{\textnormal{p}} = 1$\;\textnormal{h}
where $N_{\textnormal{bins}} = 25$ and $100$.
The displayed uncertainties include
statistical uncertainties only and
the one-sigma spreads (bands)
are the standard deviations of the
central values.
The observed distributions are
compared to a Gaussian distribution in which
only statistical errors are included
using a Kolmogorov--Smirnov test.
}
\label{fig:KS-test-x-5-8}
\end{figure}

\begin{figure}[p]
\begin{center}
\includegraphics[width=0.99 \linewidth]{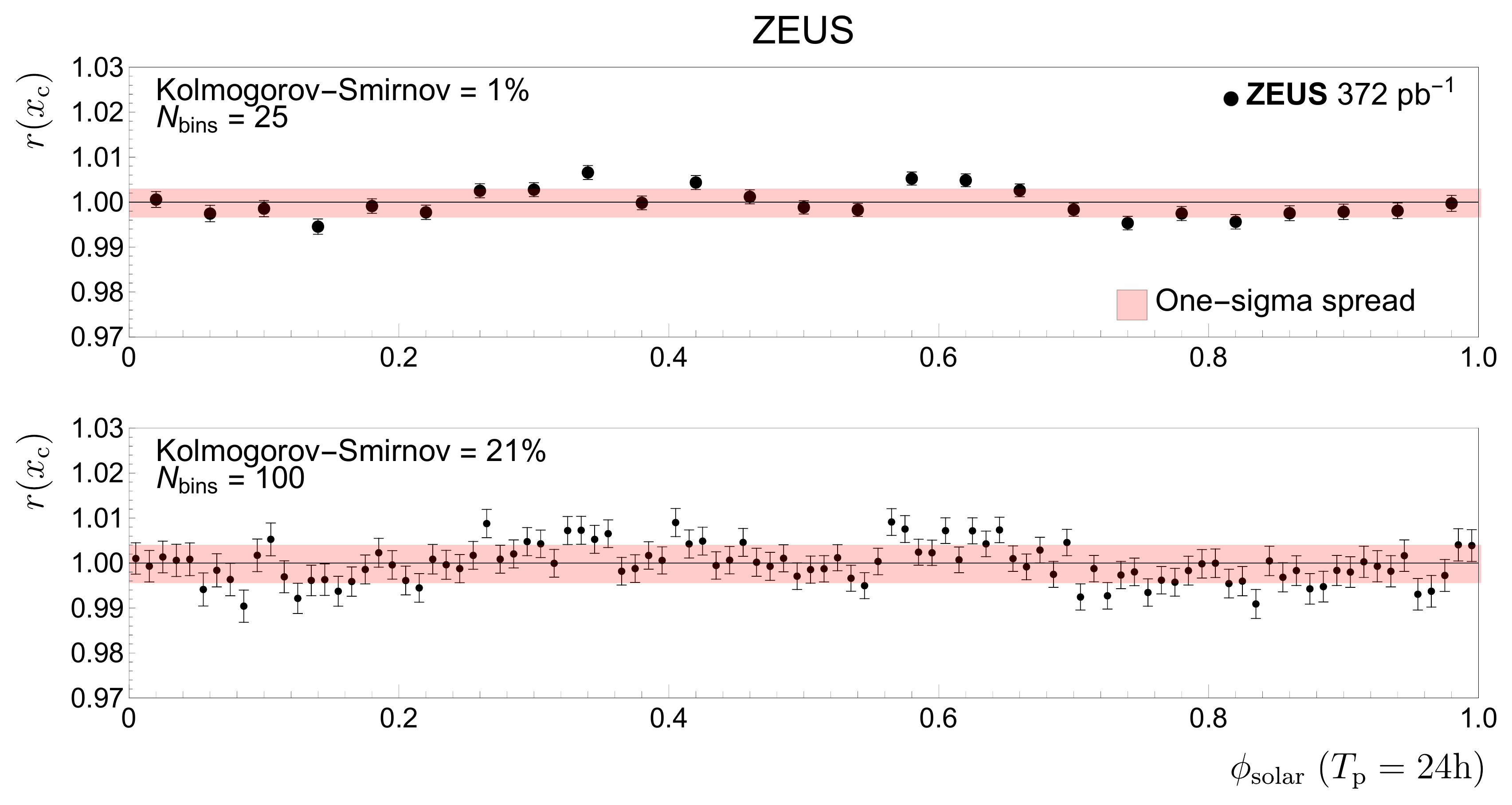}
\end{center}
\caption{
Ratio $r(x_{\textnormal{c}})$ of normalised
counts for $x_{\textnormal{c}} = 10^{-3}$
binned using the solar phase 
where $N_{\textnormal{bins}} = 25$ and $100$.
The case $N_{\textnormal{bins}} = 100$ from
Fig.~\ref{fig:DIS-0.001}(a) is repeated.
The displayed uncertainties include
statistical uncertainties only and
the one-sigma spreads (bands)
are the standard deviations of the
central values.
The observed distributions are
compared to a Gaussian distribution in which
only statistical errors are included
using a Kolmogorov--Smirnov test.
}
\label{fig:KS-solar-x-5-8}
\end{figure}

When separating the kinematic range into regions above and
below the cut $x_{\text{c}} = 10^{-3}$, the normalised counts presented 
in Fig.~\ref{fig:DIS-0.001} strongly
resemble those from $Q_{\text{c}}^2$ cuts in Fig.~\ref{fig:DIS-20}.
Similarly, in light of the discussion in Section~\ref{ssec:initial}, 
it is expected that the $T_{\text{p}}=1$\;h distributions for the ratios 
$r(x_{\text{Bj}}> x_{\text{c}} , x_{\text{Bj}}< x_{\text{c}} )$ 
given explicitly 
in Eqs.~\eqref{eq:rc} and \eqref{eq:ra5} 
will be distributed around unity and will be
dominated by statistical uncertainties.
As discussed, any potential solar and sidereal effects 
should be removed when  
binning data using a test period 
much smaller than 24 h.
This is indeed the case, as can be seen 
from the results for $\phi_{\text{test}}$ presented in 
Fig.~\ref{fig:KS-test-x-5-8}. 
In contrast to the $Q_{\text{c}}^2$ distributions for all phases, 
the solar phases for the $x_{\text{c}}$ distributions
contain more structure, as is immediately evident from Fig.~\ref{fig:KS-solar-x-5-8}.

The $N_\text{bins}=25$ configuration exhibits a 1\% KS
probability, implying the presence of systematics
that are unaccounted for.
This is also apparent 
as the one-sigma spreads, particularly for the 
smaller $N_{\text{bins}}=25$ case, 
are much wider than the 
uncertainty bars on the central values,
with larger deviation around
unity than in the test-phase 
case (see Fig.~\ref{fig:KS-test-x-5-8}). 
This is not sufficient to identify
the origin of the additional
systematic effects.
However, an estimate of this systematic contribution 
for a given $N_{\text{bins}}$ may be obtained
by calculating 
\begin{equation}
\sigma_{\text{syst}} \approx
\sqrt{\sigma^2 - \sigma^2_{\text{stat}}}\;,
\end{equation}
where $\sigma$ and $\sigma_{\text{stat}}$ 
are the standard deviation
and mean statistical uncertainty
of the points of the distribution,
respectively.

As already stressed, 
the observed systematic 
effects are not directly 
connected to large
fluctuations of the instantaneous
luminosity. They have not been
previously observed because
such small uncertainties were
negligible in all previous
ZEUS analyses.
This was confirmed by means
of a Monte Carlo study.
Inclusive DIS events 
with $Q^2 > 4 \;\text{GeV}^2$ were 
generated with DJANGOH~1.6~\cite{Schuler:1991yg} 
interfaced to ARIADNE~\cite{Pettersson:1988zu, 
Lonnblad:1988nh, Lonnblad:1989kj, Lonnblad:1992tz}. 
The CTEQ5D~\cite{Lai:1999wy} PDFs were used. 
The events were subsequently
passed through the ZEUS detector 
and trigger simulations 
based on GEANT 3~\cite{Brun:1987ma}.
The Monte Carlo events contain
leading-order QCD corrections plus 
parton showering at matrix-element 
level. Also, the $x$ in the
PDFs is no longer 
identified with $x_{\text{Bj}}$,
although this is not expected
to affect results significantly.

The parton-level calculation is based
purely on the Standard Model and does not 
introduce any direct time
dependence. The generated events 
do not possess time stamps. 
However, the detector response and the 
trigger configuration depend on the 
instantaneous luminosity 
and may change over time.
Data events are grouped into 
blocks with the same detector and 
trigger configurations
and, for each of these blocks, 
appropriate Monte Carlo 
events were generated.
In order to include time dependence 
in the Monte Carlo events and to 
simulate accurately  the impact
of the variable instantaneous luminosity 
over the course of 5\,years, the 
following procedure
was performed. For each luminosity block
containing a given number of data events, 
all time stamps were extracted and 
assigned Monte Carlo events randomly
selected from all those events generated
with the same detector and trigger status. 
The obtained Monte Carlo sample 
simulates accurately any
time dependence associated with the 
instantaneous luminosity and
the detector response. 

This study was performed on 
a subset of experimental data (namely electron 
data taken in 2006) corresponding
to an integrated luminosity of 
54.8 pb${}^{-1}$. 
It was found that
phase distributions obtained from Monte Carlo 
events are perfectly compatible with
statistical uncertainties alone: 
no evidence of any residual 
systematic contribution to the
fluctuation of the binned central 
values around unity is 
observed in these simulations.
This is also confirmed
by comparing the one-sigma spreads of 
data and Monte Carlo, with the former
being much larger than the 
latter (and the latter being in 
agreement with the statistical uncertainties alone).

This analysis shows that the residual 
systematic uncertainties observed in data 
are not modelled in existing Monte Carlo
simulations. As explained above, 
a reasonable strategy is to 
use the difference in quadrature 
between the one-sigma spreads and the 
corresponding statistical uncertainties as 
an estimate of the additional systematics.
Note that, in the present study, 
all previously known sources of systematic 
uncertainties cancel. 

As additional checks, separate studies involving 
events taken under different
trigger settings, which are constant
within each luminosity block, 
have been performed. No 
discernible impact on the presence 
of the systematic uncertainties
discussed above is observed.
\subsection{Estimated systematic uncertainties}
The binned solar $x_{\text{c}}$ analyses 
clearly indicate the presence of an unknown, 
substantial, and pervasive 
time-dependent systematic effect
with mean systematic uncertainty
$\bar\sigma_{\text{syst}}\approx 0.26\%$.
The extraction of 
systematic uncertainties for
periods 
$T_{\text{sidereal}} \approx T_\text{solar}-4$\;min
($\bar\sigma_{\text{syst}}\approx 0.18\%$) and
$2T_{\text{solar}}-T_{\text{sidereal}} \approx T_\text{solar}+4$\;min 
($\bar\sigma_{\text{syst}}\approx 0.11\%$) 
both result
in smaller systematic uncertainties
relative to the period $T_{\text{solar}}$,  
further suggesting a possible unaccounted 
for solar-periodic effect, which 
may be a consequence of the operation
of the experiment.
The systematic uncertainties on the
ratio binned in sidereal time
may therefore reflect a dilution of the observed
potential solar-periodic effect.
While, at this stage, a claim
of a genuine solar-time effect cannot be made,
there is no significant 
indication of a comparable sidereal-time effect.
This establishes a baseline for extracting constraints.
Under the assumption of a genuine solar systematic 
effect, the $2T_{\text{solar}}-T_{\text{sidereal}}$ 
distributions determine the dilution of 
the solar systematic when changing the period by $+4$\;minutes;
in fact, dilution effects for very small positive or negative time 
shifts are very similar. This method makes possible
an estimate of the
systematic uncertainties on the sidereal-time 
distributions without 
risking an absorption of potential genuine sidereal signals.

\begin{figure}[p]
\begin{center}
\includegraphics[width=0.99 \linewidth]{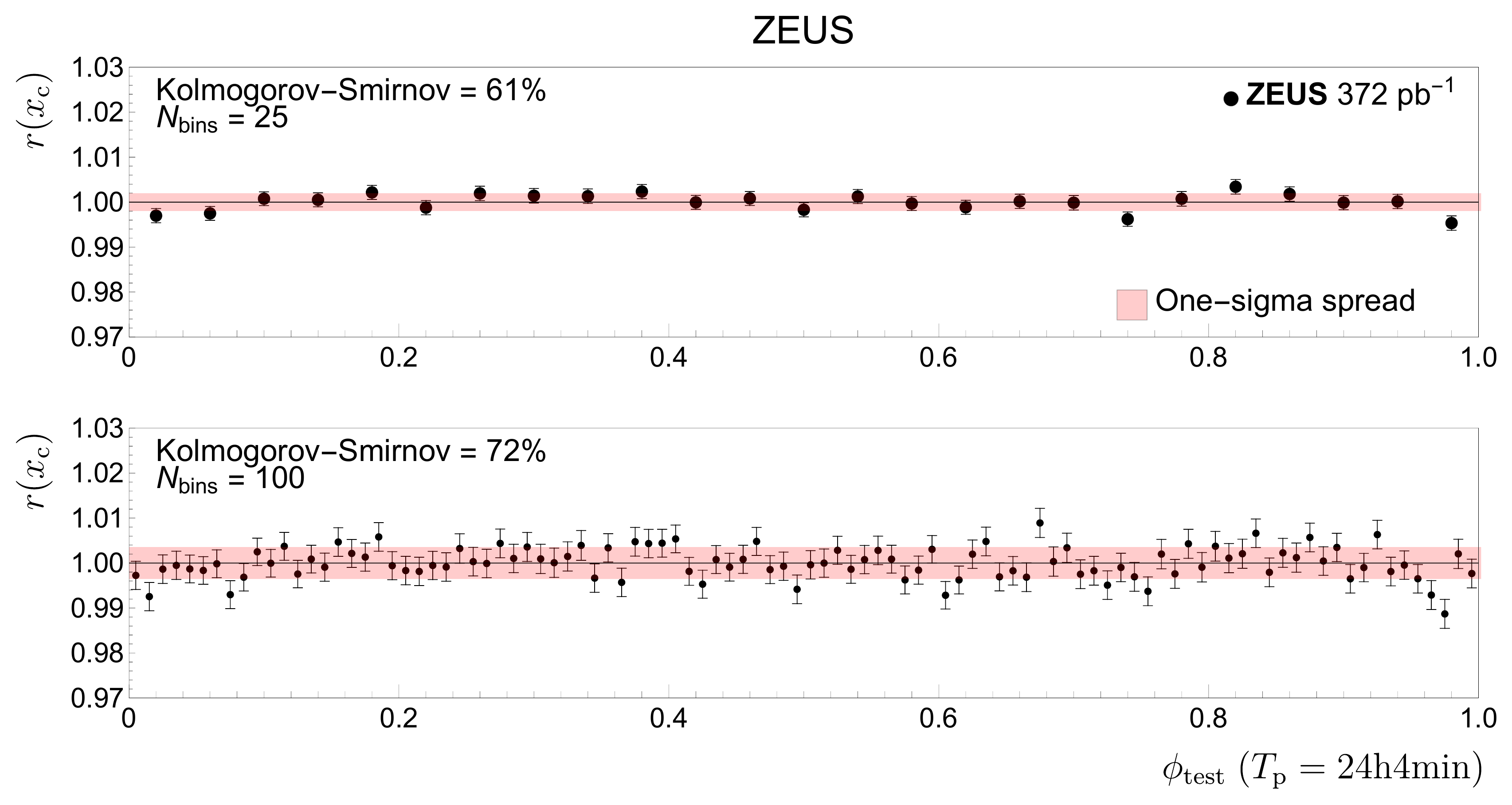}
\end{center}
\caption{
Ratio $r(x_{\textnormal{c}})$ of normalised
counts for $x_{\textnormal{c}} = 10^{-3}$
binned using the test phase with
$T_{\textnormal{p}} = 24\;\textnormal{h}\;4\textnormal{min}$
where $N_{\textnormal{bins}} = 25$ and $100$.
The displayed uncertainties include
statistical uncertainties only and
the one-sigma spreads (bands)
are the standard deviations of the
central values.
The observed distributions are
compared to a Gaussian distribution in which
only statistical errors are included
using a Kolmogorov--Smirnov test.
}
\label{fig:KS-shiftedsolar-x-5-8}
\end{figure}

In the next section, the constraints on the SME coefficients
are extracted using $N_{\text{bins}} = 100$.
For consistency, the systematic uncertainty used in the 
constraint-setting procedure is also extracted 
from the $2T_{\text{solar}}-T_{\text{sidereal}}$ distribution
with 100 bins. This distribution is presented in 
Fig.~\ref{fig:KS-shiftedsolar-x-5-8} and yields 
$\sigma_{\text{syst}}\approx 0.16\%$.
A fuller understanding of this 
systematic effect is left for 
future studies and is 
beyond the scope of the present analysis.
\section{Constraints on effective couplings}
\label{ssec:constraints}
\begin{figure}[p]
\begin{center}
\includegraphics[width=0.99 \linewidth]{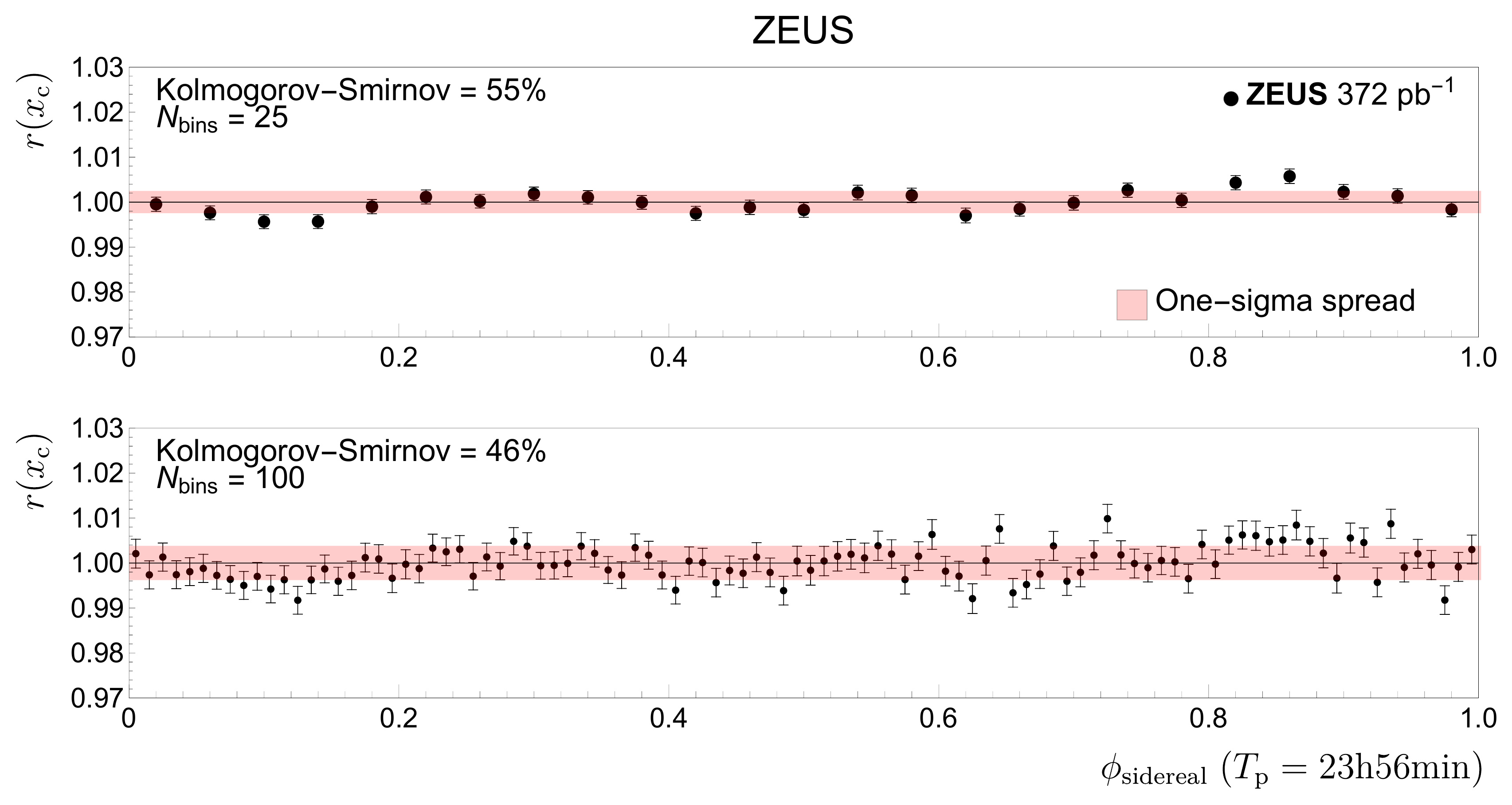}
\end{center}
\caption{
Ratio $r(x_{\textnormal{c}})$ of normalised
counts for $x_{\textnormal{c}} = 10^{-3}$
binned using the sidereal phase
where $N_{\textnormal{bins}} = 25$ and $100$.
The displayed uncertainties include
statistical uncertainties only and
the one-sigma spreads (bands)
are the standard deviations of the
central values.
The observed distributions are
compared to a Gaussian distribution in which
only statistical errors are included
using a Kolmogorov--Smirnov test.
}
\label{fig:KS-sidereal-x-5-8}
\end{figure}

%
The sidereal-phase distributions of the ratio of 
normalised counts for $x_{\text{Bj}}$ above 
and below $x_{\text{c}} = 10^{-3}$ 
are shown in Fig.~\ref{fig:KS-sidereal-x-5-8}. 
This is the main result of this paper from which 
constraints on the SME coefficients will be extracted.
Compared to the solar-phase case, 
much higher KS probabilities are observed.
To extract constraints, the following procedure was performed. 
The laboratory coefficients in the ratios 
Eqs.~\eqref{eq:rc} and \eqref{eq:ra5} were
replaced with the SCF coefficients
as explained in Section~\ref{ssec:sidsignals},
replacing one coefficient at a time.
The ratios then depend on the local sidereal angle 
$\theta_\oplus = \omega_\oplus T_\oplus$. For a given
bin $i$, the theoretical ratios $r_i^{\text{theo}}$ 
were calculated as
\begin{equation}
\label{eq:ri_theory}
r^{\text{theo}}_i
= \frac{N_{\text{bins}}}{2\pi}
\int_{\frac{2\pi(i-1)}{N_{\text{bins}}}}
^{\frac{2\pi i}{N_{\text{bins}}}} 
r(x>x_{\text{c}}, x<x_{\text{c}};\theta_\oplus)d\theta_\oplus\;,
\end{equation}
where $i=1,\ldots,N_{\text{bins}}$. 
This quantity can be compared directly 
with the experimental sidereal
ratios $r^{\text{exp}}_i$.
A $\chi^2$ function was constructed
\begin{equation}
\label{eq:chi2}
\chi^2 =
\frac{1}{\sigma_{\text{tot}}^2}\sum_{i=1}^{N_{\text{bins}}}
\left(r^{\text{exp}}_i 
- r^{\text{theo}}_i\right)^2,
\end{equation}
for each of the 42 observable SCF coefficients. 
The total uncertainty $\sigma_{\text{tot}}$ in 
Eq.~\eqref{eq:chi2} is given
by combining in quadrature
the statistical uncertainty 
($\left[\sigma_\text{stat}\right]_{N_\text{bins}=100} = 0.32\%$)
and the systematic uncertainty 
estimated in the previous section ($\sigma_\text{syst} = 0.16\%$)
taken as an additional random uncertainty at each point,
so that $\sigma_\text{tot} = 0.35\%$. 

The $\chi^2$ in the Standard Model is 
obtained by setting all the SME 
coefficients to zero 
(the theory prediction in the Standard Model 
is identically 1 in each phase bin), 
and is 113.8 for 100 degrees of freedom.
The $p$-value for this goodness-of-fit (GOF)
test is $p_\text{SM} = 0.16$, indicating
a reasonable description of 
the data by the Standard Model.
In Tables~\ref{tab:cbounds} and~\ref{tab:a5bounds}, 
lower (upper) values of each coefficient
are presented, below (above)
which the $p$-values for the same
$\chi^2$ GOF test are smaller than 0.05.
These values are indicative of 
the disfavoured ranges for which the
$c$-type coefficients
are roughly at the level 
of $10^{-4}$ for the $u$-quark coefficients
and $10^{-3}$ for the $d$- and $s$-quark coefficients.
The corresponding ranges for the $a^{(5)}$-type coefficients are
mostly at the level of $10^{-7}$\;GeV$^{-1}$
for the $u$-quark coefficients and 
$10^{-6}$\;GeV$^{-1}$ for the $d$-quark coefficients.

\begin{figure}[p]
    \begin{center}
    \includegraphics[width=0.99 \linewidth]{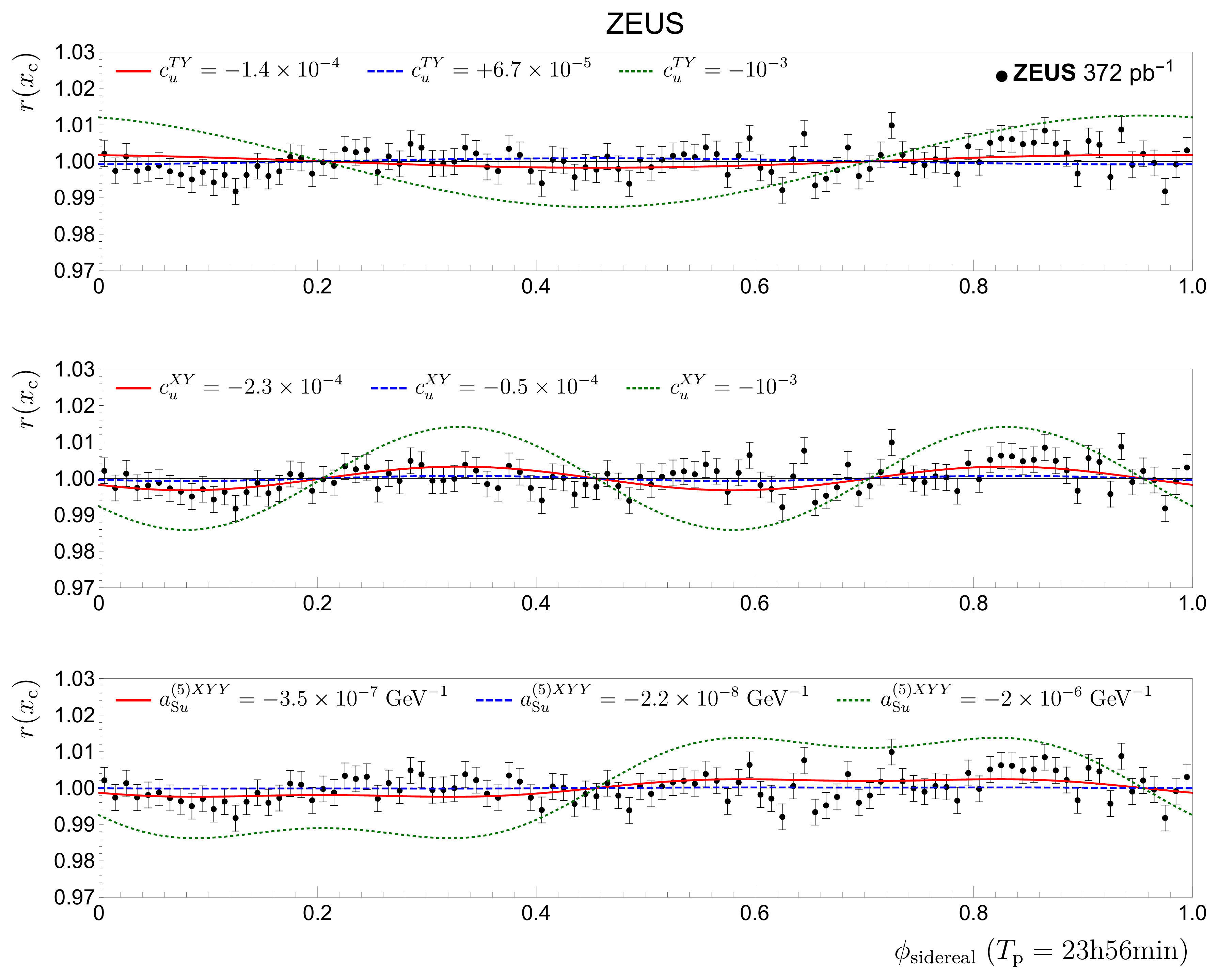}
    \end{center}
\caption{
Ratio $r(x_{\textnormal{c}})$ of normalised
counts for $x_{\textnormal{c}} = 10^{-3}$
binned using the sidereal phase
where $N_{\textnormal{bins}} = 100$.
The displayed uncertainties include
statistical uncertainties only.
In the three panels, signals corresponding to the SME coefficients 
$c_u^{TY}$, $c_u^{XX}-c_u^{YY}$ and $a_{\textnormal{S}u}^{(5)XXY}$ 
which are associated with variations up to $\omega_\oplus, 2\omega_\oplus$, 
and $3\omega_\oplus$, respectively, are displayed with
values selected at the edge of the disfavoured
ranges presented in 
Tables~\ref{tab:cbounds} and \ref{tab:a5bounds} (solid, dashed and dotted lines). In addition, signals that are excluded
by the extracted bounds are displayed with 
$c_u^{TY} = 10^{-3}$, $c_u^{XX}-c_u^{YY} = 10^{-3}$,
and $a_{\textnormal{S}u}^{(5)XXY} 
= +2\times 10^{-6}$\;\textnormal{GeV}$^{-1}$.
}
\label{fig:KS-sidereal-x-5-8-signal}
\end{figure}

Figure~\ref{fig:KS-sidereal-x-5-8-signal} 
shows the time dependence associated with 
non-vanishing $c_u^{TY}$, $c_u^{XX}-c_u^{YY}$ and 
$a_{\text{S}u}^{(5)XXY}$ coefficients. 
These three coefficients have 
been chosen because they are
examples of time dependence with 
angular frequencies $\omega_\oplus$, 
$2\omega_\oplus$ and $3\omega_\oplus$,
respectively. In each case, the data points 
are identical to those 
presented in Fig.~\ref{fig:KS-sidereal-x-5-8},
the solid and dashed curves correspond to 
selected values of 
the disfavoured regions presented in 
Tables~\ref{tab:cbounds} and \ref{tab:a5bounds}, 
and the dotted curves correspond 
to coefficients that are roughly an 
order of magnitude larger.

Comparison with existing 
constraints is informative~\cite{tables}.
For the $c$-type coefficients,
the results derived in this work
represent the first constraints 
using sidereal oscillations. For the 
$u$ and $d$ quarks, there are 
existing and considerably 
more stringent constraints
derived from interpretations
of cosmic-ray measurements~\cite{Schreck:2017isa}. 
The latter constraints rest on a number
of model-dependent assumptions, 
so that direct comparison between
those results and what is 
presented here requires caution.
The constraints on the $s$-quark coefficients 
are derived for the first time. 
%
The results for the $a^{(5)}$-type coefficients
are the first of their kind. 
One potential point 
of comparison would be the 
constraints on effective 
$a^{(5)}$-type coefficients for 
protons, as resported
in Table D11~\cite{tables}. 
Using hydrogen 1S--2S transitions,
similar combinations 
of proton constraints
are currently at the level of 
$10^{-7}$--$10^{-8}$\;GeV$^{-1}$, 
which is similar in magnitude 
to that found for $u$-quark 
$a^{(5)}$-type coefficients~\cite{Kostelecky:2015nma}.
\section{Summary and outlook}
An analysis searching for effective coefficients
parameterising Lorentz and CPT violation
in the light-quark sector has been performed
using ZEUS $e^\pm p$ DIS NC data at HERA. 
Conservative estimates of previously unknown 
time-dependent systematic uncertainties
independent of instantaneous luminosity
and trigger configuration have been made
by binning and analysing
events as a function of the 
sidereal-rotation frequency of the Earth. 
By combining binned sidereal 
statistical uncertainties with 
systematic uncertainties in quadrature,
the data are shown to be 
compatible with the Standard Model.
First constraints 
have been placed on the 24 combinations 
of the non-renormalisable, CPT-violating,
and spin-independent
$a^{(5)}$-type operators associated
with rotationally
anisotropic $u$- and $d$-quark 
effects. For the 
renormalisable CPT-preserving $c$-type coefficients,
the first constraints have been placed
on the six rotationally anisotropic
$s$-quark coefficients 
and the first experimental constraints
have been extracted on the analogous 12
$u$- and $d$-quark coefficients.
In total, 42 coefficients
have been constrained, 30 for the first time.

Partonic-physics studies of fundamental symmetries and experimental analyses searching for unconventional time-dependent signatures are
in their infancy.
A statistically significant 
indication of Lorentz or related
fundamental-symmetry violation in 
EFT would not unequivocally indicate
the presence of phenomena
outside the description of 
quantum field theory and metric
theories of gravity. Nonetheless,
such a discovery would 
indicate the presence of 
new physics, which is a strong
motivation for further
work in this area.


\section*{Acknowledgements}
\label{sec-ack}

\Zacknowledge


{
\ifzeusbst
  \ifzmcite
     \bibliographystyle{./BiBTeX/bst/l4z_default3}
  \else
     \bibliographystyle{./BiBTeX/bst/l4z_default3_nomcite}
  \fi
\fi
\ifzdrftbst
  \ifzmcite
    \bibliographystyle{./BiBTeX/bst/l4z_draft3}
  \else
    \bibliographystyle{./BiBTeX/bst/l4z_draft3_nomcite}
  \fi
\fi
\ifzbstepj
  \ifzmcite
    \bibliographystyle{./BiBTeX/bst/l4z_epj3}
  \else
    \bibliographystyle{./BiBTeX/bst/l4z_epj3_nomcite}
  \fi
\fi
\ifzbstjhep
  \ifzmcite
    \bibliographystyle{./BiBTeX/bst/l4z_jhep3}
  \else
    \bibliographystyle{./BiBTeX/bst/l4z_jhep3_nomcite}
  \fi
\fi
\ifzbstnp
  \ifzmcite
    \bibliographystyle{./BiBTeX/bst/l4z_np3}
  \else
    \bibliographystyle{./BiBTeX/bst/l4z_np3_nomcite}
  \fi
\fi
\ifzbstpl
  \ifzmcite
    \bibliographystyle{./BiBTeX/bst/l4z_pl3}
  \else
    \bibliographystyle{./BiBTeX/bst/l4z_pl3_nomcite}
  \fi
\fi
{\raggedright
\bibliography{./syn.bib,%
              ./myref.bib,%
              ./BiBTeX/bib/l4z_zeus.bib,%
              ./BiBTeX/bib/l4z_h1.bib,%
              ./BiBTeX/bib/l4z_articles.bib,%
              ./BiBTeX/bib/l4z_books.bib,%
              ./BiBTeX/bib/l4z_conferences.bib,%
              ./BiBTeX/bib/l4z_misc.bib,%
              ./BiBTeX/bib/l4z_preprints.bib}}
}
\vfill\eject


\begin{table}[p]
\begin{center}
\begin{tabular}{ccc}
\hline
Coefficient & Lower & Upper \\
\hline\hline
$c_{u}^{TX}$ & $-2.5\times 10^{-4}$ & $6.6\times 10^{-5}$ \\ 
$c_{u}^{TY}$ & $-1.7\times 10^{-4}$ & $9.8\times 10^{-5}$ \\ 
$c_{u}^{XY}$ & $-3.2\times 10^{-4}$ & $4.1\times 10^{-5}$ \\ 
$c_{u}^{XZ}$ & $-5.4\times 10^{-4}$ & $1.4\times 10^{-4}$ \\ 
$c_{u}^{YZ}$ & $-3.7\times 10^{-4}$ & $2.1\times 10^{-4}$ \\ 
$c_{u}^{XX}-c_{u}^{YY}$ & $-2.1\times 10^{-4}$ & $2.5\times 10^{-4}$ \
\\ 
& & \\ 
$c_{d}^{TX}$ & $-7.8\times 10^{-4}$ & $2.0\times 10^{-4}$ \\ 
$c_{d}^{TY}$ & $-5.2\times 10^{-4}$ & $3.0\times 10^{-4}$ \\ 
$c_{d}^{XY}$ & $-1.6\times 10^{-3}$ & $2.0\times 10^{-4}$ \\ 
$c_{d}^{XZ}$ & $-2.7\times 10^{-3}$ & $7.0\times 10^{-4}$ \\ 
$c_{d}^{YZ}$ & $-1.8\times 10^{-3}$ & $1.0\times 10^{-3}$ \\ 
$c_{d}^{XX}-c_{d}^{YY}$ & $-1.0\times 10^{-3}$ & $1.2\times 10^{-3}$ \
\\ 
& & \\ 
$c_{s}^{TX}$ & $-9.6\times 10^{-4}$ & $2.5\times 10^{-4}$ \\ 
$c_{s}^{TY}$ & $-6.4\times 10^{-4}$ & $3.7\times 10^{-4}$ \\ 
$c_{s}^{XY}$ & $-2.6\times 10^{-3}$ & $3.3\times 10^{-4}$ \\ 
$c_{s}^{XZ}$ & $-4.4\times 10^{-3}$ & $1.2\times 10^{-3}$ \\ 
$c_{s}^{YZ}$ & $-3.0\times 10^{-3}$ & $1.7\times 10^{-3}$ \\ 
$c_{s}^{XX}-c_{s}^{YY}$ & $-1.7\times 10^{-3}$ & $2.0\times 10^{-3}$ \
\\ 
\hline\hline
\end{tabular}
\end{center}
\caption{Lower (upper) values 
of the $c$-type coefficients
below (above)
which the $p$-values for the $\chi^2$
GOF test are smaller than 0.05.
}
\label{tab:cbounds}
\end{table}

\begin{table}[p]
\begin{center}
\begin{tabular}{ccc}
\hline
Coefficient & Lower\; (GeV${}^{-1}$)&  Upper\; (GeV${}^{-1}$) \\
\hline \hline 
$a_{\text{S}u}^{(5)TXX}-a_{\text{S}u}^{(5)TYY}$ & $-5.1\times 10^{-7}$ & $4.3\times 10^{-7}$ \\ 
$a_{\text{S}u}^{(5)XXZ}-a_{\text{S}u}^{(5)YYZ}$ & $-1.7\times 10^{-6}$ & $2.0\times 10^{-6}$ \\ 
$a_{\text{S}u}^{(5)TXY}$ & $-8.3\times 10^{-8}$ & $6.5\times 10^{-7}$ \\ 
$a_{\text{S}u}^{(5)TXZ}$ & $-2.9\times 10^{-7}$ & $1.1\times 10^{-6}$ \\ 
$a_{\text{S}u}^{(5)TYZ}$ & $-4.3\times 10^{-7}$ & $7.4\times 10^{-7}$ \\ 
$a_{\text{S}u}^{(5)XXX}$ & $-3.9\times 10^{-7}$ & $1.2\times 10^{-7}$ \\ 
$a_{\text{S}u}^{(5)XXY}$ & $-2.3\times 10^{-7}$ & $1.8\times 10^{-7}$ \\ 
$a_{\text{S}u}^{(5)XYY}$ & $-4.6\times 10^{-7}$ & $9.2\times 10^{-8}$ \\ 
$a_{\text{S}u}^{(5)XYZ}$ & $-2.6\times 10^{-6}$ & $3.3\times 10^{-7}$ \\ 
$a_{\text{S}u}^{(5)XZZ}$ & $-5.4\times 10^{-7}$ & $1.4\times 10^{-7}$ \\ 
$a_{\text{S}u}^{(5)YYY}$ & $-2.9\times 10^{-7}$ & $1.5\times 10^{-7}$ \\ 
$a_{\text{S}u}^{(5)YZZ}$ & $-3.6\times 10^{-7}$ & $2.1\times 10^{-7}$ \\ 
& & \\ 
$a_{\text{S}d}^{(5)TXX}-a_{\text{S}d}^{(5)TYY}$ & $-7.3\times 10^{-6}$ & $6.1\times 10^{-6}$ \\ 
$a_{\text{S}d}^{(5)XXZ}-a_{\text{S}d}^{(5)YYZ}$ & $-2.4\times 10^{-5}$ & $2.8\times 10^{-5}$ \\ 
$a_{\text{S}d}^{(5)TXY}$ & $-1.2\times 10^{-6}$ & $9.4\times 10^{-6}$ \\ 
$a_{\text{S}d}^{(5)TXZ}$ & $-4.1\times 10^{-6}$ & $1.6\times 10^{-5}$ \\ 
$a_{\text{S}d}^{(5)TYZ}$ & $-6.1\times 10^{-6}$ & $1.1\times 10^{-5}$ \\ 
$a_{\text{S}d}^{(5)XXX}$ & $-5.7\times 10^{-6}$ & $1.7\times 10^{-6}$ \\ 
$a_{\text{S}d}^{(5)XXY}$ & $-3.4\times 10^{-6}$ & $2.7\times 10^{-6}$ \\ 
$a_{\text{S}d}^{(5)XYY}$ & $-6.8\times 10^{-6}$ & $1.3\times 10^{-6}$ \\ 
$a_{\text{S}d}^{(5)XYZ}$ & $-3.7\times 10^{-5}$ & $4.6\times 10^{-6}$ \\ 
$a_{\text{S}d}^{(5)XZZ}$ & $-8.1\times 10^{-6}$ & $2.1\times 10^{-6}$ \\ 
$a_{\text{S}d}^{(5)YYY}$ & $-4.3\times 10^{-6}$ & $2.3\times 10^{-6}$ \\ 
$a_{\text{S}d}^{(5)YZZ}$ & $-5.4\times 10^{-6}$ & $3.1\times 10^{-6}$ \\ 
\hline\hline
\end{tabular}
\end{center}
\caption{Lower (upper) values 
of the $a^{(5)}$-type coefficients
below (above)
which the $p$-values for the $\chi^2$
GOF test are smaller than 0.05.
}
\label{tab:a5bounds}
\end{table}

%
%
\end{document}